\documentclass[aps,physrev,preprint,groupedaddress]{revtex4-2}
\usepackage{graphicx}
\usepackage{xcolor}
\usepackage{subcaption}
\usepackage{amsmath}
\usepackage{float}

\renewcommand{\vec}[1]{\mathbf{#1}}
\newcommand{\phip}{\phi_{\text{p}}}

\definecolor{myblue}{rgb}{0.9,0.0,1.0}

\begin{document}


\title{Influence of near-field effect on magnetic hysteresis in magneto-active elastomers}


\author{Pawan Patel$^{1,2}$, Dirk Romeis$^{*\,1}$ and Marina Saphiannikova$^{1,2}$}
\affiliation{$^{1}$Division Theory of Polymers, Leibniz Institute of Polymer Research Dresden, 01069 Dresden, Germany}
\affiliation{$^{2}$Faculty of Mechanical Science and Engineering, Technical University Dresden, 01062 Dresden, Germany}
\email{Corresponding author: romeis@ipfdd.de}

\date{\today}

\begin{abstract}

 Magneto-active elastomers (MAEs) are polymer composites consisting of magnetic microparticles embedded in an elastomeric matrix. These materials exhibit strong magneto-mechanical coupling under external magnetic fields, resulting in tunable stiffness, reversible shape changes, and nonlinear magnetic responses. This study presents a multiscale theoretical framework to investigate the origin of magnetic hysteresis in MAEs, with emphasis on the evolution of the internal microstructure during magnetization and demagnetization.
 The total energy of the system is formulated as the sum of magnetic and micromechanical contributions, while macroscopic deformation of a cylindrical MAE sample is fully constrained. Particle interactions are modeled first via pure dipole–dipole interactions and then extended to include higher-order near-field effects at close particle separations. The results show that hysteresis in MAEs with magnetically soft particles primarily arises from trapped microstructural rearrangements, leading to distinct particle configurations under increasing and decreasing magnetic fields. Parametric studies demonstrate that particle volume fraction, sample aspect ratio, and matrix stiffness strongly influence the microstructure evolution and the width of resulting hysteresis loops. The proposed framework provides a solid foundation for modeling magnetic hysteresis, which is essential for the design and optimization of MAEs in practical applications.

\end{abstract}


\maketitle

\section{Introduction}
Magneto-active elastomers (MAEs) are soft composites made by dispersing magnetic microparticles within an elastomeric matrix, such as silicone rubber (PDMS), natural rubber, or polyurethane. They exhibit field-tunable mechanical and physical properties. When exposed to a magnetic field, the composites can change stiffness, shape, and surface morphology in a reversible way. The magneto-mechanical coupling enables MAEs for medical devices \cite{Sun2024, Liu2024, Makarova_2016, Makarova_2017}, soft robotics \cite{Reiche2024, Han_2024} and active surfaces \cite{Kriegl_2024, Strauss2024, Sanchez_2019}. MAEs can be classified as hard or soft magnetic materials depending on the properties of their particle inclusions. MAEs containing magnetically hard inclusions, such as NdFeB or ferrites, exhibit high coercivity and remanent magnetization \cite{Hermann2020,MORENO_MATEOS2023,Ye2021}. In contrast, MAEs with magnetically soft inclusions, such as carbonyl iron or iron-silicon particles, show negligible remanence and respond reversibly to external magnetic fields \cite{Moreno2021,GARCIAGONZALEZ2021108796,Charles2020}. The focus of the current study are MAEs with magnetically soft particle inclusions. When these materials are subjected to an external magnetic field, magnetization is induced inside the micron-sized particles. This magnetization increases with the strength of the applied magnetic field, as has been observed in numerous experiments \cite{Borin_2020,Becker_2020,Maas_2016}. In addition, the induced magnetization is measured upon decreasing the magnetic field in order to examine its field-dependent reversibility. A mismatch between the magnetic response in increasing and decreasing fields results in a phenomenon known as magnetic hysteresis. Such hysteretic behavior has been reported for MAEs in experimental studies \cite{KRAUTZ2017,Borin_2022}. 

With the aim of investigating the origins of hysteresis, Biller et al. \cite{Biller2014} performed a theoretical analysis of particle pairs in magnetorheological elastomers. This analysis revealed that bistability and hysteresis appear only within a restricted domain of the material parameter space (elastic modulus, saturation magnetization, initial interparticle distance).  Within this domain the system exhibits an abrupt first‑order transition between a separated configuration and when particles are in close contact, leading to a hysteresis loop whose width is controlled by the ratio of magnetic to elastic energy and by the initial interparticle distance. In a follow up study, Biller et al. \cite{Biller2015} modeled the elastic response of the matrix using a Mooney‑Rivlin‐type hyperelastic energy. 
They demonstrated that, in dense MAEs, short‑range multipolar magnetic interactions significantly exceed the point‑dipole approximation. The resulting coupled magneto‑elastic energy landscape exhibits two competing minima, providing a mesoscopic mechanism for the macroscopic magneto-mechanical hysteresis observed experimentally. The idea to model hysteresis in MAEs via the bistability of a particle pair was further investigated and refined \cite{Vaganov2020, Biller2024}. 

To study the magnetic properties of MAEs, Zubarev et al. \cite{Zubarev2016} proposed a hierarchical model of chain formation, in which the magnetic attraction between linear chains competes with elastic resistance upon their approach. In this framework, the aggregation threshold for chain growth during magnetic field increase differs from the disintegration threshold upon field decrease, which directly results in hysteresis in magnetization and susceptibility curves \cite{Zubarev2016}. A similar observation was reported in experiments by Glavan et al. \cite{GLAVAN2023}. These experiments suggest that the hysteresis originates from differences in the spatial arrangement of particles under increasing and decreasing magnetic fields. It is worth mentioning that Stepanov et al. \cite{Zubarev_2025} modeled the mechanical hysteresis of MAE subjected to an external magnetic field. The model fits the experimental data accurately.  

\begin{figure}[ht]
\centering
\begin{subfigure}{0.18\textwidth}
    \centering
    \includegraphics[width=\linewidth]{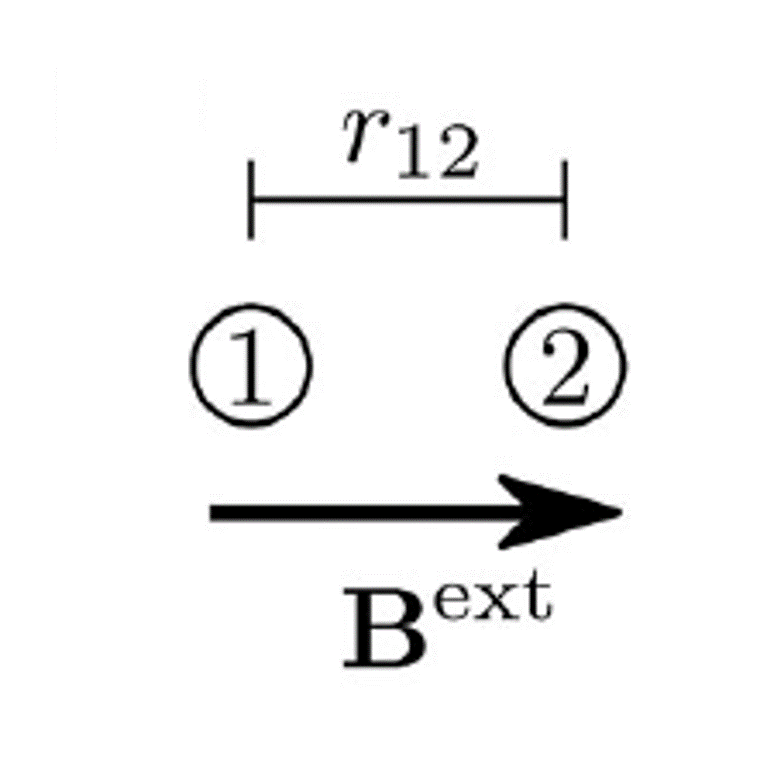}
    \caption{}
\end{subfigure}
\hfill
\begin{subfigure}{0.18\textwidth}
    \centering
    \includegraphics[width=\linewidth]{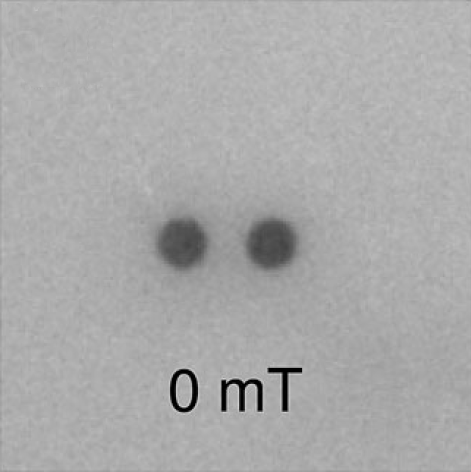}
    \caption{}
\end{subfigure}
\hfill
\begin{subfigure}{0.18\textwidth}
    \centering
    \includegraphics[width=\linewidth]{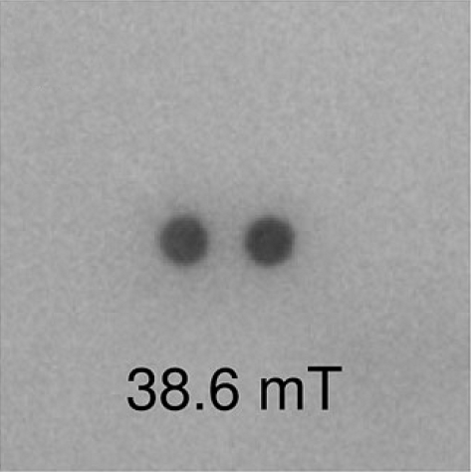}
    \caption{}
\end{subfigure}
\hfill
\begin{subfigure}{0.18\textwidth}
    \centering
    \includegraphics[width=\linewidth]{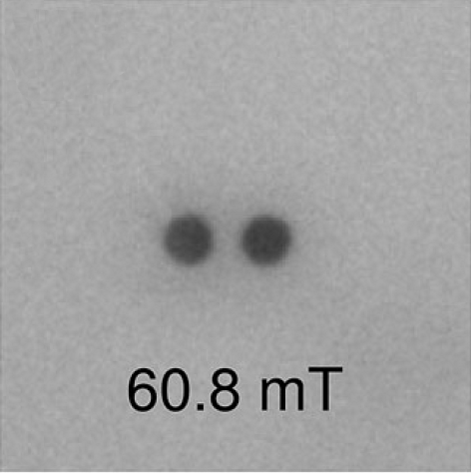}
    \caption{}
\end{subfigure}
\hfill
\begin{subfigure}{0.18\textwidth}
    \centering
    \includegraphics[width=\linewidth]{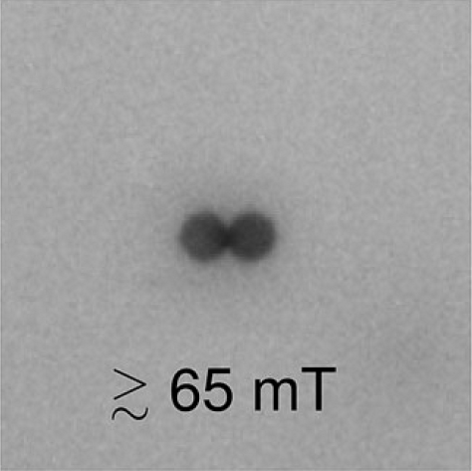}
    \caption{}
\end{subfigure}
  
\caption{The magneto-mechanical collapse in a system of two magnetic particles upon increase of the magnetic field. Adapted from Ref.~\cite{Puljiz2018}. Copyright 2018 Royal Society of Chemistry.}
\label{fgr: Fig2}
\end{figure}

Building on the concept of two competing minima \cite{Biller2015}, Puljiz et al. \cite{Puljiz2018} experimentally demonstrated a pronounced hysteresis in the particle separation‑field relationship (see Fig.~\ref{fgr: Fig2}). The experiments showed that two paramagnetic nickel inclusions embedded in a soft PDMS matrix undergo “virtual touching” at a critical field (~65 mT). When the field is removed, the particles separated and return to their initial position. Analytical linear‑elastic calculations, dipole‑spring reductions and finite‑element simulations confirmed that the hysteresis arises from the coexistence of two local minima in the total energy and from kinetic trapping in one of these minima, depending on the history of the applied magnetic field \cite{Puljiz2018}. 

Another approach to modeling magnetic hysteresis involves the use of numerical simulations, with Finite Element Analysis (FEA) being the most popular tool. Two complementary modeling strategies are pursued. Microscopic approaches resolve the explicit particle‑matrix microstructure using a representative volume element (RVE). The RVE captures the geometry of each inclusion and its spatial arrangement in the finite element mesh. The total Lagrangian finite element formulation solves the coupled magneto‑mechanical boundary‑value problem over each element thereby yielding the local magnetization and stress field. The volume average of local magnetization and local stress yields the effective macroscopic behavior \cite{Kalina2020,Metsch_2021}. Macroscopic phenomenological models treat the composite as a homogeneous continuum and incorporate magneto‑elastic coupling terms calibrated either from experimental data or from up-scaled microscopic simulations, yielding effective material parameters \cite{Kalina2020}. Both strategies rely on thermodynamically consistent constitutive formulations derived from a free‑energy density. To model hysteresis within the framework of generalized standard materials, Gebhart et al. \cite{Gebhart2025} introduced a dissipation potential into the total energy density \cite{Gebhart2025}.

The current study builds on the mean-field approach proposed in our group \cite{Romeis_2021}. This approach enables us to consider macroscopic samples containing a large number of magnetic particles, with $\phi$ being their total volume fraction. In the absence of an external magnetic field, the particles are uniformly distributed over the matrix, and the material is fully isotropic. When the applied field increases, the particles start aligning themselves along the direction of the magnetic field and form anisotropic columnar structures. In our mean-field approach, we do not consider discrete particle distributions, but smear the particle positions inside the columnar structures, as schematically represented in Fig.~\ref{fgr: Fig1}.  
The fundamental quantity to account for the microstructure evolution is the local volume fraction of particles in the columns, $\phip$.
In the absence of the magnetic field $\phip$ is equal to $\phi$ and it becomes considerably larger than $\phi$ in strong magnetic fields. It is obvious that appearance of columnar structures should enhance the sample magnetization, since it is measured along the long axis of the columns. To describe this enhancement in comparison to the isotropic MAE samples with a stiff matrix, we developed models (\cite{Roghani2025,Patel_2026}) which are able to predict the magnetization curve when the applied field is increased. A good agreement with experimental results is observed for the magnetization and differential magnetic susceptibility
in the MAE samples with different matrix stiffness. 
\begin{figure}[ht]
\centering
\includegraphics[width=15cm, height=8.2cm]{Schematic_MAE_Expt.png}
\caption{Left: MAE disc between the poles of a VSM device \cite{Roghani2025}, with an external magnetic field $H_0$ applied along the symmetry axis $X$ (height $h$, diameter $D$). Right: Schematic of microstructure evolution in MAEs: increasing local field $H$ induces densely packed columnar structures with local particle volume fraction $\phi_p \gg \phi$.}
\label{fgr: Fig1}
\end{figure}

In the present study, we further develop our mean-field approach to account for the hysteretic behavior of magnetization curves. The general theoretical framework is introduced in Section 2. The main novelty is the inclusion of higher-order interactions beyond the dipole-dipole approximation, considered for both linear and saturating magnetization. In Section 3, we first present analytical results for the linear case, allowing the main trends to be identified explicitly. We then address the practically relevant case of saturating magnetization, which requires a numerical solution based on a self-consistent iterative procedure. The paper concludes with a brief discussion of the main parameters governing magnetic hysteresis.


\section{Theory}
The first step to capture hysteresis is to compute the total energy of the system. It comprises three contributions: magnetic, mechanical, and micromechanical energy, as described in previous studies \cite{Roghani2023, Roghani2025}. The model effectively accounts for both macro and microscale influences on the properties of the MAE. In the present study, cylindrical samples, as depicted in Fig.~\ref{fgr: Fig1}, are considered. The aspect ratio is varied to study the shape effect. The macroscopic deformation of the sample is completely restricted, and hence the contribution of mechanical energy can be neglected. The total energy equation governing an MAE sample is formulated as follows \cite{Roghani2025}:
\begin{equation}
\label{eq_tot_energy}
  \Psi_{\text{MAE}}(\vec{H}) = \Psi_{\text{mag}}(\vec{H}) + \Psi_{\text{mic}}(\vec{H}),
\end{equation}
where $\Psi_{\text{mag}}$ denotes the magnetic energy density, $\Psi_{\text{mic}}$ represents the micromechanical energy density associated with the microstructure evolution, and $\vec{H}$ is the internal magnetic field. Similar to the previous studies \cite{Roghani2025,Patel_2026}, the inhomogeneities of the internal field in a cylindrical sample are neglected, and it is assumed that the sample is homogeneously magnetized under the uniform external field $\vec{H}_0$. With the experimental setup of Fig.~\ref{fgr: Fig1}, the magnetization $\vec{M}$ is induced parallel to the external field $\vec{H}_0$. Thus the magnetic energy can be described based on the scalar values of the internal magnetic field $H$, external magnetic field $H_0$, and magnetization $M$. The generalized expression for magnetic energy is given in \cite{Roghani2023}:
\begin{equation}
\label{eq_mag_energy}
  \Psi_{\text{mag}} = \mu_0 \, \phi \left( -\int_{0}^{H} M dH + \frac{1}{2} \, M(H - H_0) \right),
\end{equation}
where $\mu_0$ is the permeability of vacuum. The internal magnetic field $H$ is described as a superposition of external magnetic field and the demagnetizing field $H_d$: 
\begin{equation}
\label{eq_H_Hd}
  H = H_0 + H_d.
\end{equation}
Two cases are considered. The first case is based on the pure dipole-dipole interaction model \cite{Roghani2023}. In the second case, the contribution of Near Field Effect (NFE) is added to the pure dipole-dipole model. This accounts for the higher-order interactions between the particles. 
The demagnetizing field for the pure dipole-dipole model is defined as:
\begin{equation}
\label{eq_Hd_DP}
    H_d = -\left(\nu_{\text{p}} - \phi\, f_{\text{macro}} - f_{\text{micro}} \right) M,
\end{equation}
and it is modified to account for the NFE as follows:
\begin{equation}
\label{eq_Hd_NFE}
    H_d = -\left(\nu_{\text{p}} - \phi\, f_{\text{macro}} - f_{\text{micro}} - f_{\text{NFE}} \right) M.
\end{equation}
In both expressions, $\nu_{\text{p}}=1/3$ is the the self-demagnetization factor of a spherical inclusion, $f_{\text{macro}}$ represents the effect of macroscopic shape on the internal magnetic field, $f_{\text{micro}}$ incorporates the microstructural effects and $f_{\text{NFE}}$ denotes the NFE. The macroscopic shape factor $f_{\text{macro}}$ is a function of the demagnetizing factor $N_{\parallel}$ for a cylinder magnetized homogeneously along the symmetry axis:
\begin{equation}
\label{eq_fmacro}
    f_{\text{macro}} = \frac{1}{3} - N_{\parallel}.
\end{equation}
 The demagnetizing factor is approximated by \cite{Sato_1989}:
\begin{equation}
\label{eq_demag_shape}
    N_{\parallel} = \frac{1}{\frac{4\gamma}{\sqrt{\pi}} + 1}.
\end{equation}
where $\gamma = \frac{h}{D}$ is the aspect ratio of the sample (see Fig.~\ref{fgr: Fig1}). The microstructure factor
$f_{\text{micro}}$ is proportional to the difference between the local volume fraction of particles in columns and the total volume fraction \cite{Chougale_2022} :
\begin{equation}
\label{eq_fmicro}
    f_{\text{micro}} = \frac{\phip - \phi}{3}.
\end{equation}

The pure dipole-dipole interaction model is unable to accurately capture the interaction of particles when they are in close proximity, especially when they are almost touching each other \cite{Biller2014}. As outlined by Biller et al. \cite{Biller2014}, NFE interactions are represented in the form of infinite series expansions called multipole expansions. Since purely analytical treatment of the extensive multipole expansion is not feasible, particularly for a further minimization procedure, the interpolation formula is constructed. This interpolation formula is developed by using truncated numerical evaluations of the multipole expansion (e.g., calculations involving around 100 multipole terms) as a highly accurate benchmark. Using this approach, the term accounting for the NFE is derived in this study, as detailed in Appendix:
\begin{equation}
\label{eq_fNFE}
    f_{\text{NFE}} = \frac{\phip}{9}.
\end{equation}
To simplify the notation and treat both pure dipole interactions and the case including NFE in a unified manner, we rewrite the NFE parameter as
$f_{\text{NFE}}=(\alpha-1/3)\phip$, where $\alpha=1/3$ corresponds to pure dipole interactions and $\alpha=4/9$ accounts for NFE. Furthermore, 
microstructure evolution is described in terms of the local particle enrichment
$\Delta\Phi = \phi_{\text{p}}-\phi$, defined relative to the average volume fraction $\phi$. The demagnetizing field is then denoted as:
\begin{equation}
\label{eq_Hd_unified}
    H_d = -\left(\nu_{\text{p}}- \phi\, f_{\text{macro}} - \alpha\Delta\Phi-\left(\alpha-\frac{1}{3}\right)\phi~ \right) M.
\end{equation}
This concludes the formulation of the magnetic energy term with and without NFE. The micromechanical energy contribution is described next, representing the elastic penalty associated with particle rearrangement within the elastomeric matrix \cite{Roghani2023}:
\begin{equation}
\label{eq_mic_energy}
    \Psi_{\text{mic}} = \frac{1}{2} k_m G_{iso} \left( f_{\text{micro}} - f_{\text{micro}}^{0} \right)^{2}.
\end{equation}
Here $k_m$ is the micromechanical constant, $G_{iso}$ is the elastic modulus of the matrix, and $f_{\text{micro}}^{0}$ indicates the initial state of the microstructure before an external magnetic field is applied. Since the particles are uniformly and randomly distributed in the beginning, $\phi_p = \phi$ and $f_{\text{micro}}^{0} = 0$ in the absence of a magnetic field. The total energy is computed as the sum of magnetic energy and micromechanical energy, both being a function of $\phi_p$. 

\section{Results}

In the following the results of the theory are presented and discussed. Two types of magnetization behavior are considered, linear and saturation magnetization. Although in real systems there is always saturation, the regime of small magnetic fields can be well described by linear magnetization behavior. Additionally, and in contrast to saturation magnetization, the linear case can be solved analytically. This provides direct insights to the impact of individual parameters and their relations to each other. 

\subsection{Analytic solution for linear magnetization behavior}

Assuming a linear relation between the internal magnetic field $H$ and the induced magnetization $M$, $M=\chi H$ with magnetic susceptibility $\chi$,
the magnetic energy in Eq.\ \ref{eq_mag_energy} reduces to:
\begin{equation}
\label{eq_linmag}
 \Psi_{\text{mag}} = -\frac{\mu_0 \phi M H_0}{2}~.
\end{equation}
The self-consistent solution for $M$, obtained by substituting Eq.\ \ref{eq_Hd_unified} into Eq.\ \ref{eq_H_Hd}, reads:
\begin{equation}
\label{eq_Mlin_solution}
M = \frac{H_0}{A-\alpha\Delta\Phi}~.
\end{equation}
Here we introduce the parameter
\begin{equation}
\label{eq_A_para}
A=\frac{1}{\chi}+\nu_{\text{p}}-\phi f_{macro} -\left(\alpha-\frac{1}{3}\right)\phi~,
\end{equation}
that collects all invariant sample characteristics.
The total energy $\Psi_{\text{tot}}$ (Eq.\ \ref{eq_tot_energy}) of an initially isotropic sample can then be expressed in reduced form:
\begin{equation}
\label{eq_tot_energy_linear}
\tilde{\Psi}(\Delta\Phi)=\frac{18\Psi_{\text{tot}}}{k_{\text{m}}G_{\text{iso}}}=\Delta\Phi^2-\frac{K\phi}{A-\alpha \Delta\Phi}~,
\end{equation}
where the parameter $K$ quantifies the magnetic energy relative to the elastic energy:
\begin{equation}
\label{eq_K_para}
K=\frac{9\mu_0H_0^{2}}{k_{\text{m}}G_{\text{iso}}}~.
\end{equation}
In the following, the reduced total energy is analyzed in detail under the assumption that the local particle density $\phi_{\text{p}}$ does not exceed a prescribed maximum value $\phi_{\text{max}}$. Accordingly, the local particle enrichment $\Delta\Phi$ is restricted to the interval $[0,\Delta\Phi_{\text{max}}]$, where $\Delta\Phi_{\text{max}}=\phi_{\text{max}}-\phi$.

\subsubsection{The scenario of increasing field strength}

The value of $\Delta\Phi$ which minimizes the total energy in Eq.\ \ref{eq_tot_energy_linear} is denoted by $\Delta\Phi^{*}=\phi^{*}_{\text{p}}-\phi$ in the following. It defines the equilibrium particle enrichment within the columnar microstructure for a given applied field $H_0$, or equivalently the parameter $K$. In the absence of an external field ($H_0=0$, $K=0$), this minimum occurs at $\Delta\Phi^{*}=0$. With increasing field strength, $\Delta\Phi^{*}$ increases continuously. At sufficiently large values of $\Delta\Phi$, the reduced energy $\tilde{\Psi}$ develops a maximum, as illustrated by the red and blue curves in Fig.\ \ref{fig_Utot_phistar}(a). Beyond this maximum, the energy may even fall below its value at $\Delta\Phi^{*}$ (blue curve). However, the energy barrier separating these two states prevents a spontaneous transition, such that the particle distribution remains trapped at $\Delta\Phi^{*}$. We note that the minimum at large $\Delta\Phi$ corresponds to a boundary value $\Delta\Phi_{\text{max}}$. This boundary is indicated by the vertical dashed line in Fig.\ \ref{fig_Utot_phistar}(a). A transition into the densely clustered state (approximately represented by the green curve) occurs only when both the first and the second derivatives of Eq.\ \ref{eq_tot_energy_linear} with respect to $\Delta\Phi$ vanish simultaneously. This saddle-point condition defines a critical value of the parameter $K$:
\begin{equation}
\label{eq_K_upjump}
K^{*\uparrow} = \frac{8A^3}{27\alpha^2\phi}~,
\end{equation}
at which the critical particle enrichment $\Delta\Phi^{*\uparrow}=\phi^{*\uparrow}_{\text{p}} - \phi$ jumps to the boundary value $\Delta\Phi_{\text{max}}$. The corresponding local volume fraction within the columnar structures is given by:
\begin{equation}
\label{eq_phi_upjump}
\phi^{*\uparrow}_{\text{p}} = \phi + \frac{A}{3\alpha}~.
\end{equation}

\subsubsection{The scenario of decreasing field strength}

We consider a very high initial magnetic field strength, such that the particle distribution corresponds to the densely clustered state (purple curve in Fig.\ \ref{fig_Utot_phistar}(a). Upon decreasing the magnetic field, i.e., reducing $K$, a local minimum emerges at $\Delta\Phi^{*}$ in addition to the global minimum at $\Delta\Phi_{\text{max}}$ (green and blue curves). With further decrease of the field, the minimum at $\Delta\Phi^{*}$ becomes energetically favorable and eventually turns into the global minimum. Despite this change in the energy landscape, the system remains trapped in the densely clustered state at $\Delta\Phi_{\text{max}}$. Only when the magnetic field is reduced to the point where Eq.~\ref{eq_tot_energy_linear} develops a local maximum at $\Delta\Phi_{\text{max}}$, does the clustered configuration become unstable and disperse. The particle enrichment within the columnar structures is then governed by $\Delta\Phi^{*}$, which subsequently decreases continuously with the external field and vanishes in the zero-field limit, $\Delta\Phi^{*}=0$.
The condition for Eq.~\ref{eq_tot_energy_linear} to exhibit a local maximum at $\Delta\Phi_{\text{max}}$ reads:
\begin{equation}
\label{eq_K_downjump}
 K^{*\downarrow} = \frac{2\Delta\Phi_{\text{max}}\left(A-\alpha \Delta\Phi_{\text{max}}\right)^2}{\alpha\phi}~.
\end{equation}
Interestingly, for $A=\alpha \Delta\Phi_{\text{max}}$ the critical value becomes zero. Consequently, the system jumps back to the fully dispersed state only in the limit of a completely vanishing applied field, $H_0 \to 0$. Using the definitions of $A$ and $\Delta\Phi_{\text{max}}$, the condition $A\leq\alpha \Delta\Phi_{\text{max}}$ can be rewritten as:
\begin{equation}
\label{eq_phi_treshold}
\alpha\phi_{\text{max}}\geq\frac{1}{\chi}+\frac{1}{3}-\phi \left(f_{\text{macro}} - \frac{1}{3}\right)~.
\end{equation}
In order to minimize the right-hand side of Eq.~\ref{eq_phi_treshold}, we consider the extreme limit $\chi\to\infty$ and $f_{\text{macro}}\to 1/3$, corresponding to a prolate sample shape with infinite aspect ratio. In this case, Eq.~\ref{eq_phi_treshold} reduces to
\begin{equation}
\phi_{\text{max}}\geq \frac{1}{3\alpha}.
\end{equation}
Within the dipole approximation, this would require $\phi_{\text{max}}\geq 1$, while inclusion of NFE yields $\phi_{\text{max}}\geq 0.75$. Both values exceed physically accessible volume fractions. In practice, the maximum achievable packing fraction is expected to be close to random close packing of spheres, $\phi_{\text{rcp}}\approx 0.65$. \cite{larson1999structure} Hence, we choose $\phi_{\text{max}}=0.65$ in the following. The minimum of $\tilde{\Psi}$ in Eq.~\ref{eq_tot_energy_linear} determines the particle volume fraction inside the columnar structures after the discontinuous transition. Solving $\partial_{\Delta\Phi}\tilde{\Psi}=0$ at $K=K^{*\downarrow}$ yields the physically relevant minimum:
\begin{equation}
\label{}
\phi^{*\downarrow}_{\text{p}}=\phi+\frac{A}{\alpha}-\frac{\phi_{\text{max}}-\phi}{2}-\sqrt{\frac{A(\phi_{\text{max}}-\phi)}{\alpha}-\frac{3(\phi_{\text{max}}-\phi)^{2}}{4}}~.
\end{equation}
Thus, the particle volume fraction inside the columnar structures 'jumps' from $\phi_{\text{max}}$ to $\phi^{*\downarrow}_{\text{p}}$.

\subsubsection{Upper critical value where no 'jump' is possible}

For certain parameter combinations, it may occur that $K^{*\downarrow} \geq K^{*\uparrow}$. In this case, no hysteresis is observed: both the local volume fraction $\phi_{\text{p}}$ and the magnetization vary smoothly and reversibly along the same branch upon increasing and decreasing the external magnetic field. The absence of a discontinuous transition is associated with the emergence of a saddle point at $\Delta\Phi_{\text{max}}$, or beyond. Imposing that both the first and second derivatives of $\tilde{\Psi}$ in Eq.~\ref{eq_tot_energy_linear} vanish at $\Delta\Phi = \Delta\Phi_{\text{max}}$ yields the condition $A \geq 3\alpha \Delta\Phi_{\text{max}}$. Using the definition of $A$, this inequality defines a threshold volume fraction $\phi^{**}$ given by:
\begin{equation}
\label{}
\phi^{**} = \frac{3\alpha\phi_{\text{max}}-\frac{1}{\chi}-\frac{1}{3}}{2\alpha-f_{\text{macro}}+\frac{1}{3}}~.
\end{equation}
This implies that for total particle volume fractions $\phi > \phi^{**}$, our theory within the present linear approximation predicts the absence of hysteresis.
\begin{figure}[ht]
 \centering
    \includegraphics[width=0.49\textwidth]{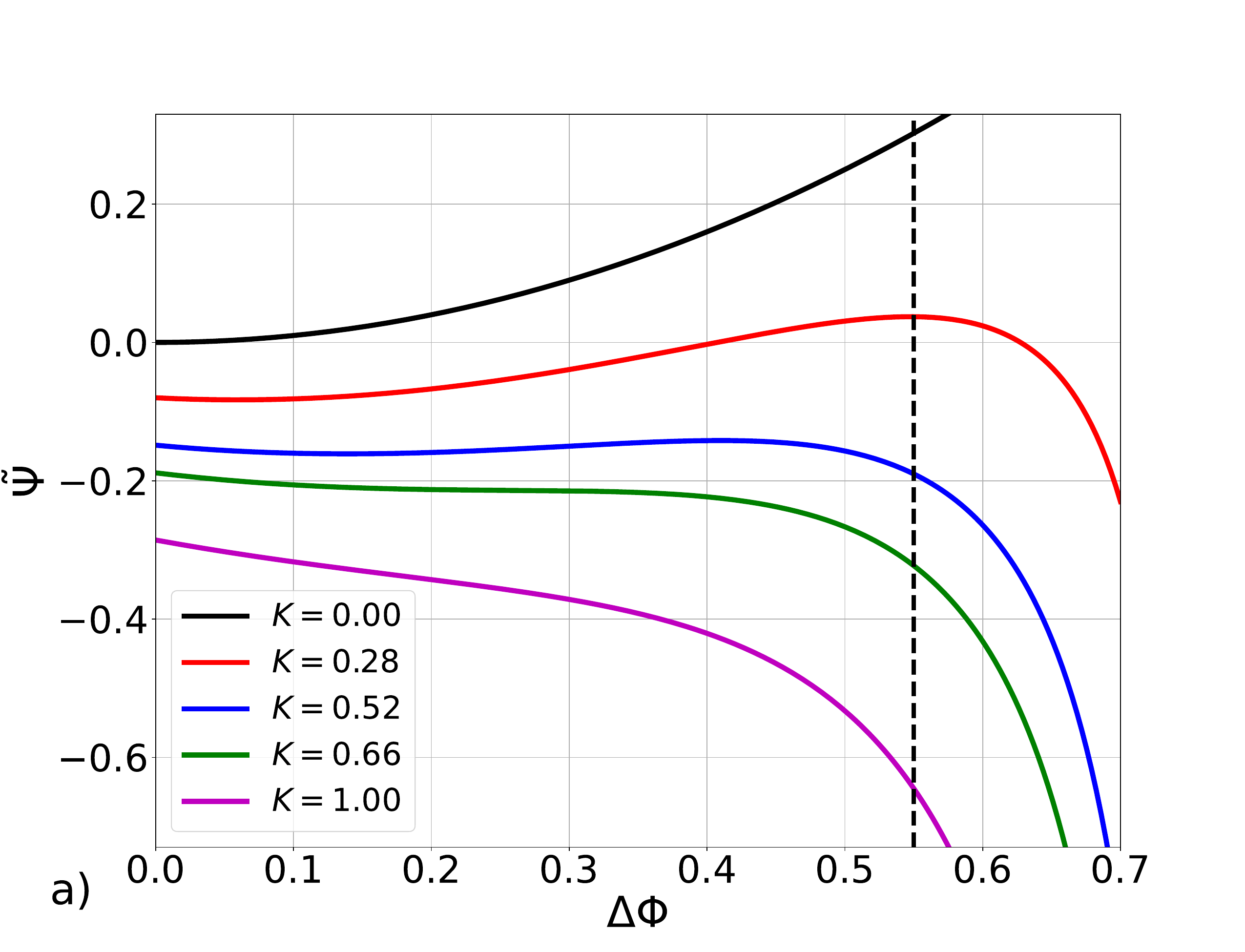}\includegraphics[width=0.49\textwidth]{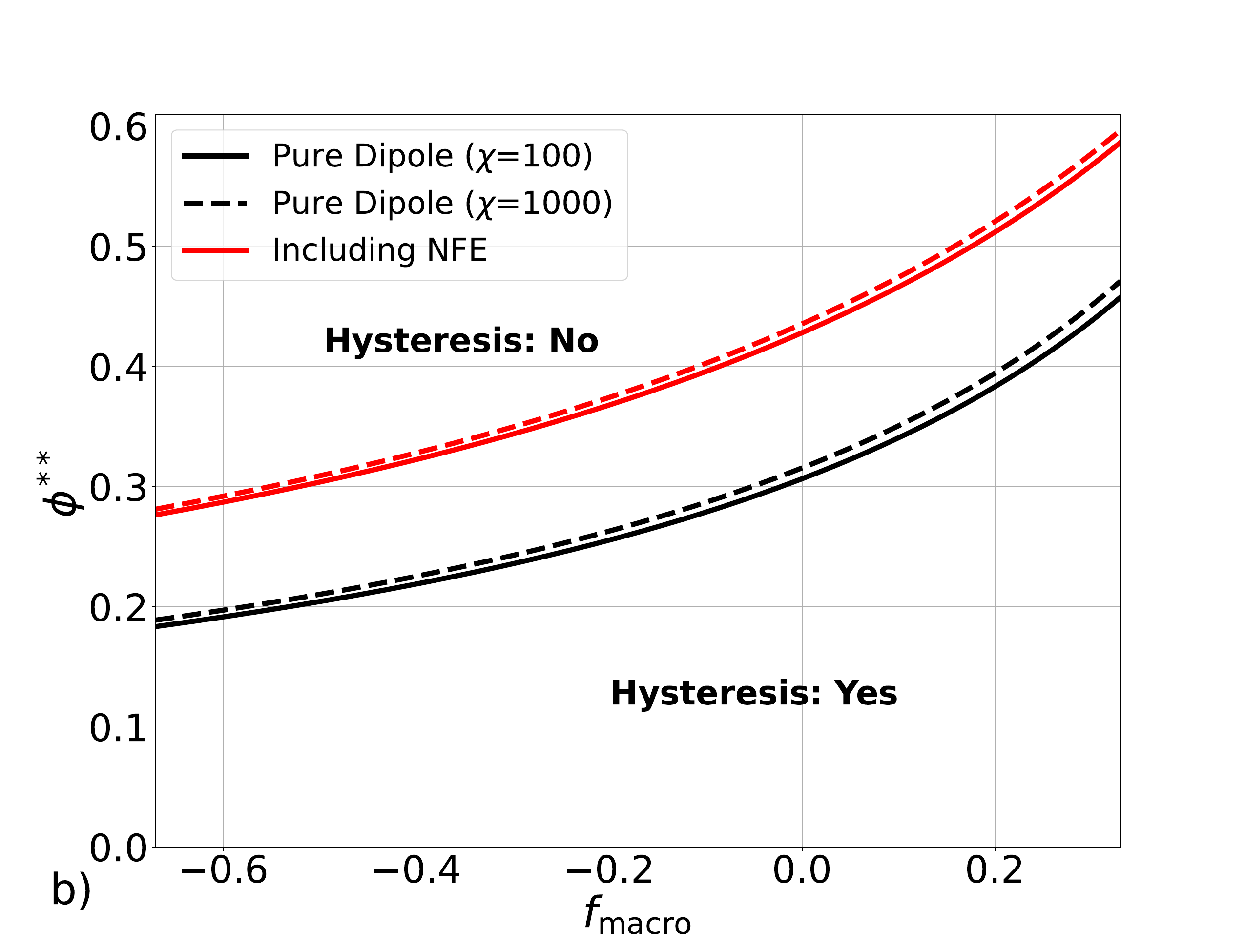}
 \caption{\label{fig_Utot_phistar} (a) Example energy landscape from Eq.\ \ref{eq_tot_energy_linear} with $A=0.3$ and $\alpha=4/9$ (including NFE). The total particle volume fraction $\phi$ can take any value in $(0,\phi_{\text{max}})$. Here, the densely clustered state $\Delta\Phi_{\text{max}}=0.55$ (marked by a dashed vertical line) corresponds to $\phi=0.10$ for $\phi_{\text{max}}=0.65$. (b) Threshold volume fraction $\phi^{**}$ as a function of  $f_{\text{macro}}$ for two susceptibilities $\chi$, shown with (red) and without (black) NFE. For $\phi \ge \phi^{**}$, the model predicts a fully reversible response without hysteresis.}
\end{figure}
The threshold volume fraction $\phi^{**}$ is shown in Fig.~\ref{fig_Utot_phistar}(b) as a function of the macroscopic shape factor $f_{\text{macro}}$. The curves are evaluated for two different values of the susceptibility $\chi$, with $\phi_{\text{max}}=0.65$. Including NFE significantly increases $\phi^{}$ (red curves). Moreover, the threshold increases with increasing susceptibility $\chi$ (dashed curves) and with increasing aspect ratio of the sample (i.e., larger $f_{\text{macro}}$). For small $\chi$ and strongly oblate sample geometries, however, the threshold can be as low as $\phi^{**}\approx 0.25$–$0.3$, even when NFE is taken into account.
Such total particle volume fractions are experimentally accessible in MAEs, suggesting that the predicted disappearance of hysteresis at sufficiently large $\phi$ could be verified experimentally.
This behavior is physically intuitive: the discontinuous transition between densely clustered and dispersed particle distributions becomes increasingly unlikely with growing particle concentration. At large $\phi$, the dispersed state is already relatively dense, so that the difference between the two configurations progressively diminishes. In this regime, the formation of distinct microstructures loses significance, as the particles are distributed throughout the sample and strong spatial contrast can no longer develop.

\subsubsection{Analysis of the linear magnetization regime}

In the following, we focus on the regime $\phi < \phi^{**}$, where a discontinuous transition and, consequently, magnetic hysteresis are predicted. Upon increasing the external field, the transition occurs at $K^{*\uparrow}$: the local particle volume fraction inside the columnar structures jumps from $\phi_{\text{p}}^{\uparrow} = \phi + \frac{A}{3\alpha}$ to $\phi_{\text{max}}$. Conversely, upon decreasing the field, the transition at $K^{*\downarrow}$ is characterized by a jump from $\phi_{\text{max}}$ to $\phi_{\text{p}}^{*\downarrow}$. Using the definition of $K$ in Eq.~\ref{eq_K_para}, the corresponding critical field strengths $H_0^{*\uparrow}$ and $H_0^{*\downarrow}$ can be determined.
\begin{figure}[ht]
 \centering
    \includegraphics[width=0.49\textwidth]{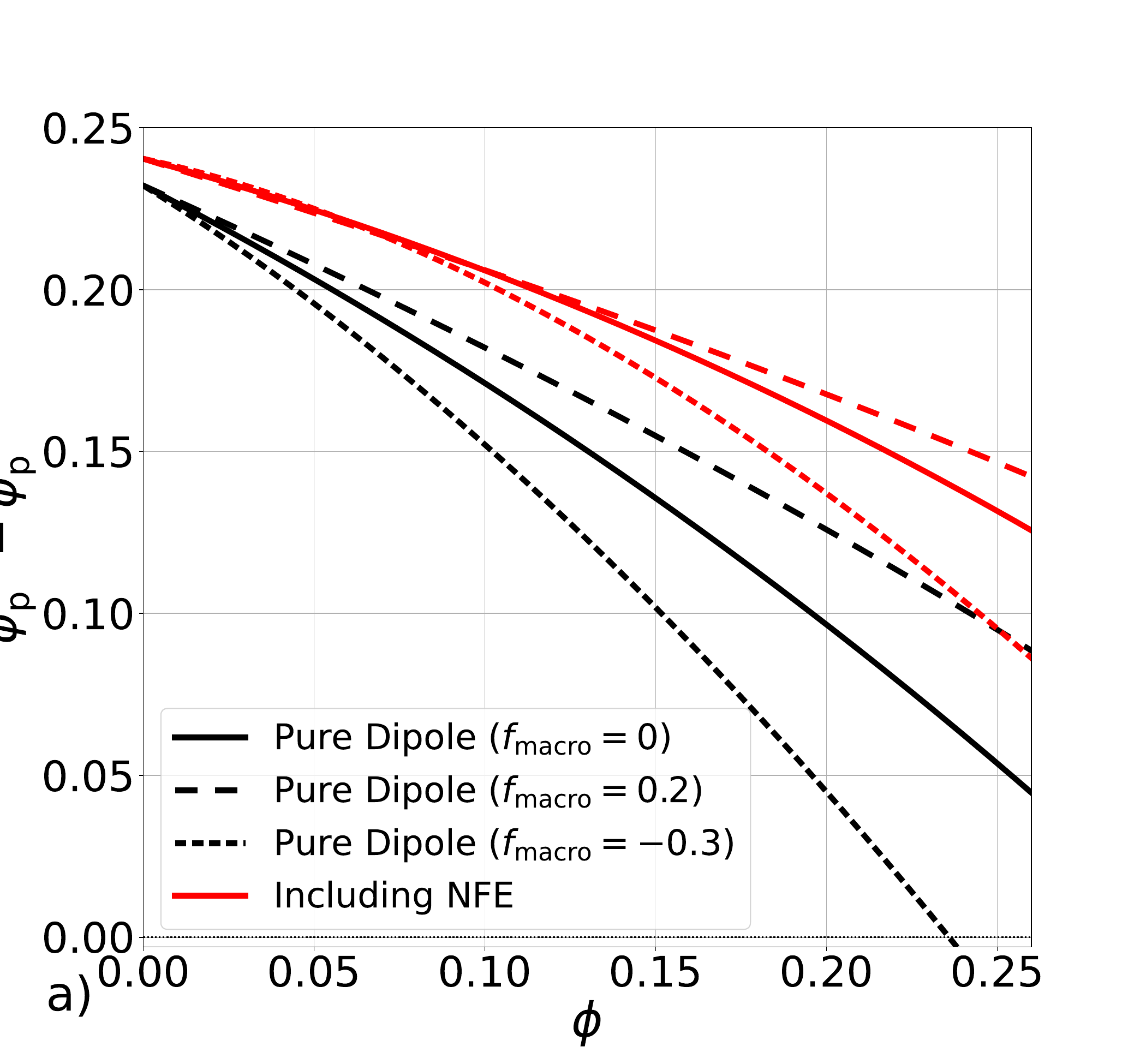}\includegraphics[width=0.49\textwidth]{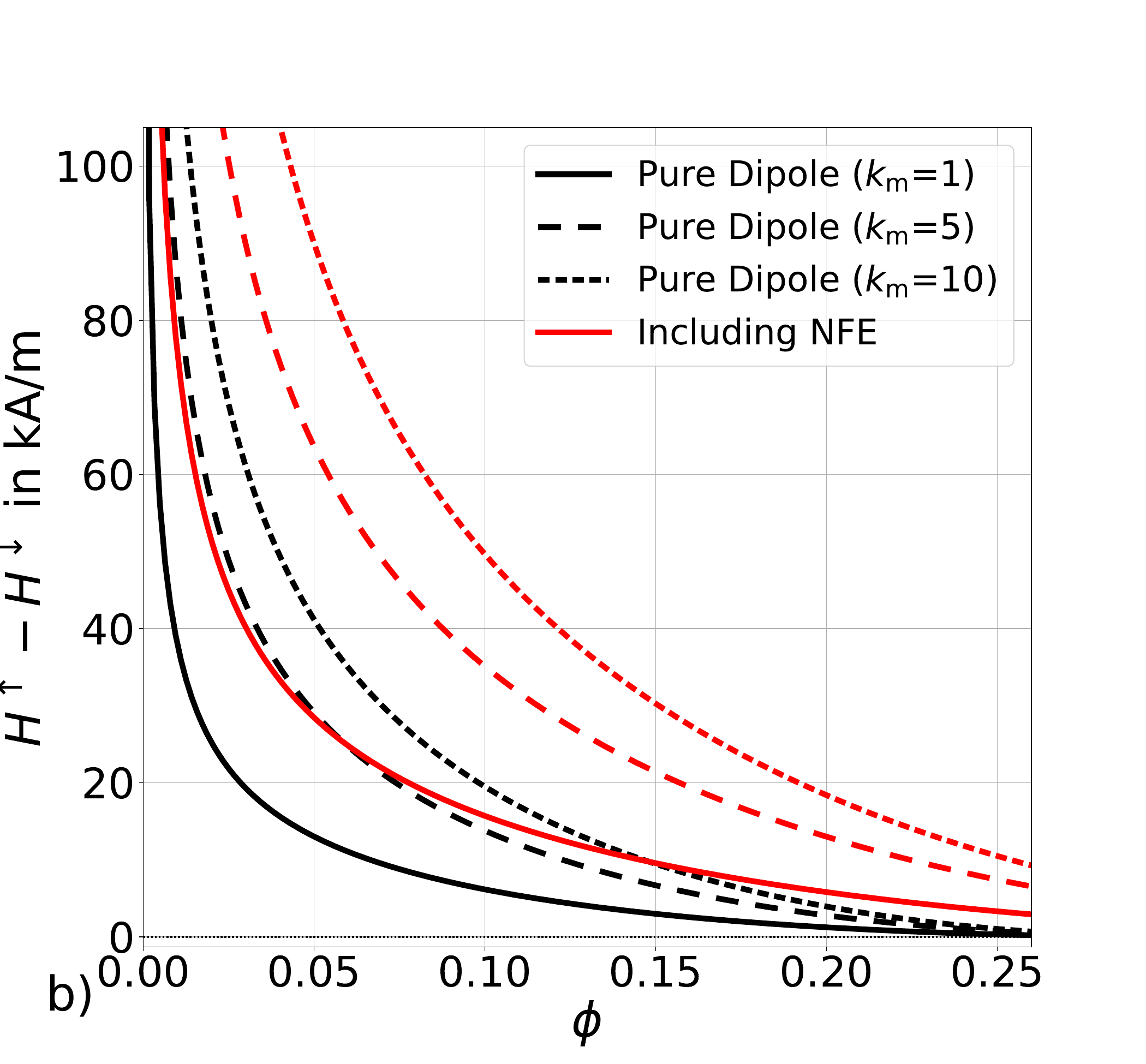}
 \caption{\label{fig_crit_phi_and_H} (a) The difference between the critical volume fractions inside columnar structures in increasing ($\phi^{*\uparrow}_{\text{p}}$) and decreasing ($\phi^{*\downarrow}_{\text{p}}$) external field, just before, resp.\ after, the densely clustered state. The parameter settings here are $\phi_{\text{max}}=0.65$ and $\chi=100$. (b) The difference between the critical external field strength when the 'jump' occurs in increasing ($H_0^{*\uparrow}$) and decreasing ($H_0^{*\downarrow}$) field. The parameter settings here are $\phi_{\text{max}}=0.65$, $\chi=100$, $f_{\text{macro}}=0$ and $G_{\text{iso}}=30\,\text{kPa}$. The red lines include NFE and the black lines are for pure dipole interactions.}
\end{figure}
The difference between $H_0^{*\uparrow}$ and $H_0^{*\downarrow}$ defines the horizontal width of the hysteresis loop, while the difference between $\phi_{\text{p}}^{*\uparrow}$ and $\phi_{\text{p}}^{*\downarrow}$ quantifies the vertical hysteresis shift. In Fig.~\ref{fig_crit_phi_and_H}(a), we plot $\phi_{\text{p}}^{*\uparrow}-\phi_{\text{p}}^{*\downarrow}$ as a function of the total particle volume fraction $\phi$ for different cases, with and without NFE. Notably, for pure dipole interactions and $f_{\text{macro}}=-0.3$ (oblate sample geometry), the hysteresis vanishes for $\phi \gtrsim 0.23$.
Fig.~\ref{fig_crit_phi_and_H}(b) shows the difference in the critical field strengths, $H_0^{*\uparrow}-H_0^{*\downarrow}$. The macroscopic shape factor $f_{\text{macro}}$ has only a minor influence on this horizontal hysteresis width. Therefore, the right panel presents results for different values of the micromechanical parameter $k_{\text{m}}$, which characterizes the mechanical contribution to the response. Notably, while $\phi_{\text{p}}^{*\uparrow}-\phi_{\text{p}}^{*\downarrow}$ exhibits a pronounced dependence on $f_{\text{macro}}$, it is entirely independent of $k_{\text{m}}$. In contrast, the field hysteresis width $H_0^{*\uparrow}-H_0^{*\downarrow}$ is strongly affected by $k_{\text{m}}$. As expected, inclusion of NFE leads to an overall increase of the hysteresis loop.
Furthermore, the theory predicts that decreasing $\phi$, combined with increasing $f_{\text{macro}}$ and $k_{\text{m}}$, enhances the hysteresis effect.

These trends merit further discussion. The influence of $f_{\text{macro}}$ is intuitive: increasing the sample aspect ratio reduces the demagnetization factor, thereby strengthening the overall magnetization response of the sample and, consequently, the hysteresis. The roles of $\phi$ and $k_{\text{m}}$, however, are more subtle in this context. With increasing $k_{\text{m}}$, the matrix effectively becomes stiffer, i.e., larger elastic energy is required to rearrange the particles. At first glance, one might therefore expect that hysteresis should become less pronounced, since particle mobility is reduced and the formation of densely clustered microstructures is hindered. However, an increase in $k_{\text{m}}$ also increases the critical field strength $H_0^{*\uparrow}$ (see Eq.~\ref{eq_K_para}), thereby facilitating the dispersion of particles. At the same time, it also increases $H_0^{*\downarrow}$, meaning that the formation of clustered structures is likewise promoted. Consequently, both critical field strengths increase with $k_{\text{m}}$, but their separation $H_0^{*\uparrow}-H_0^{*\downarrow}$ increases as well. This results in wider hysteresis loops shifted towards higher field strengths. Within the present linear magnetization approximation, this increase is not bounded. In practice, however, magnetic saturation will limit the effect, suggesting the existence of an optimal range of $k_{\text{m}}$ for which hysteresis is most pronounced. This will be demonstrated clearly in the subsection B which considers the case of saturation magnetization.

Concerning the dependence on $\phi$, we have already discussed that hysteresis is expected to weaken at very high particle concentrations, since the contrast between dispersed and clustered states becomes marginal. However, in the opposite limit $\phi \to 0$, hysteresis should likewise disappear, as cluster formation becomes increasingly unlikely when particles rarely encounter each other. Such finite-size and correlation effects are neglected within the present mean-field framework, and the dilute limit $\phi \to 0$ cannot be described realistically here. Nevertheless, the expected vanishing of hysteresis at both very small and very large $\phi$ suggests that more refined models should predict an optimal intermediate range of particle volume fractions for which hysteresis effects are most pronounced and experimentally detectable.
\subsection{Saturation magnetization}
\begin{figure}[htbp]
    \centering

    \begin{subfigure}{0.4\textwidth}
        \centering
        \includegraphics[width=\linewidth]{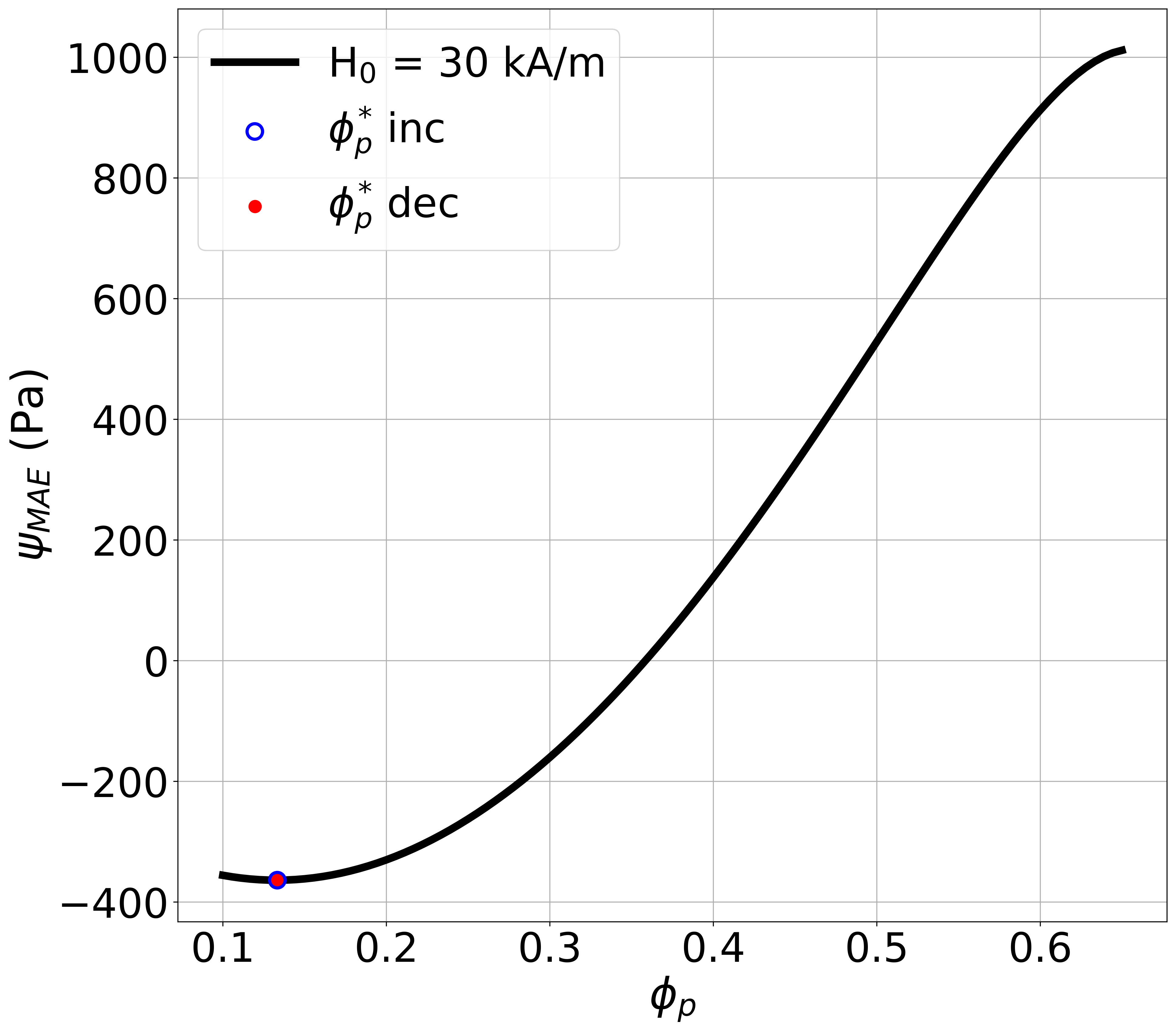}
        \caption{}
        \label{fig:a}
    \end{subfigure}
    \hfill
    \begin{subfigure}{0.4\textwidth}
        \centering
        \includegraphics[width=\linewidth]{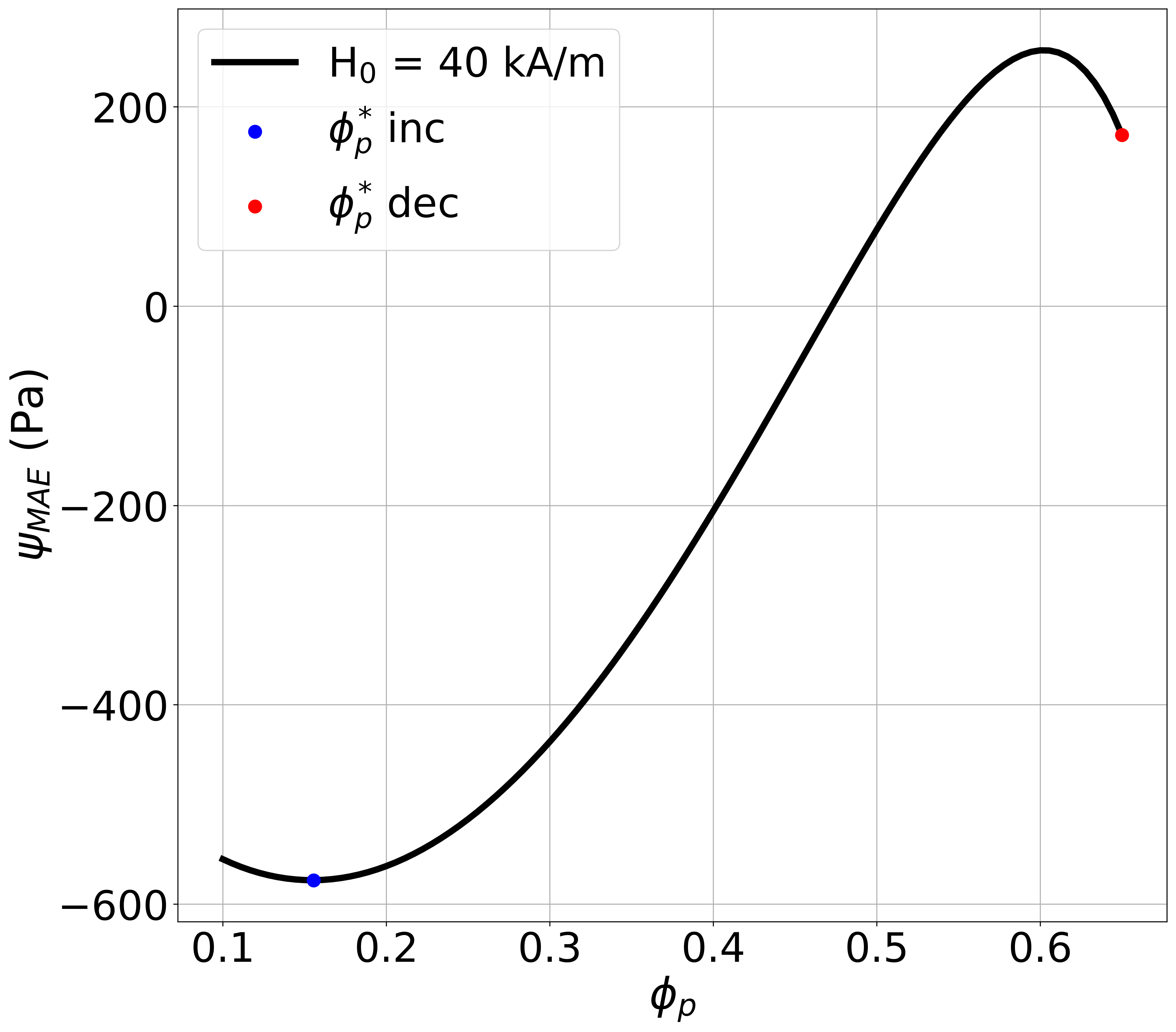}
        \caption{}
        \label{fig:b}
    \end{subfigure}

    \vspace{0.15cm}

    \begin{subfigure}{0.4\textwidth}
        \centering
        \includegraphics[width=\linewidth]{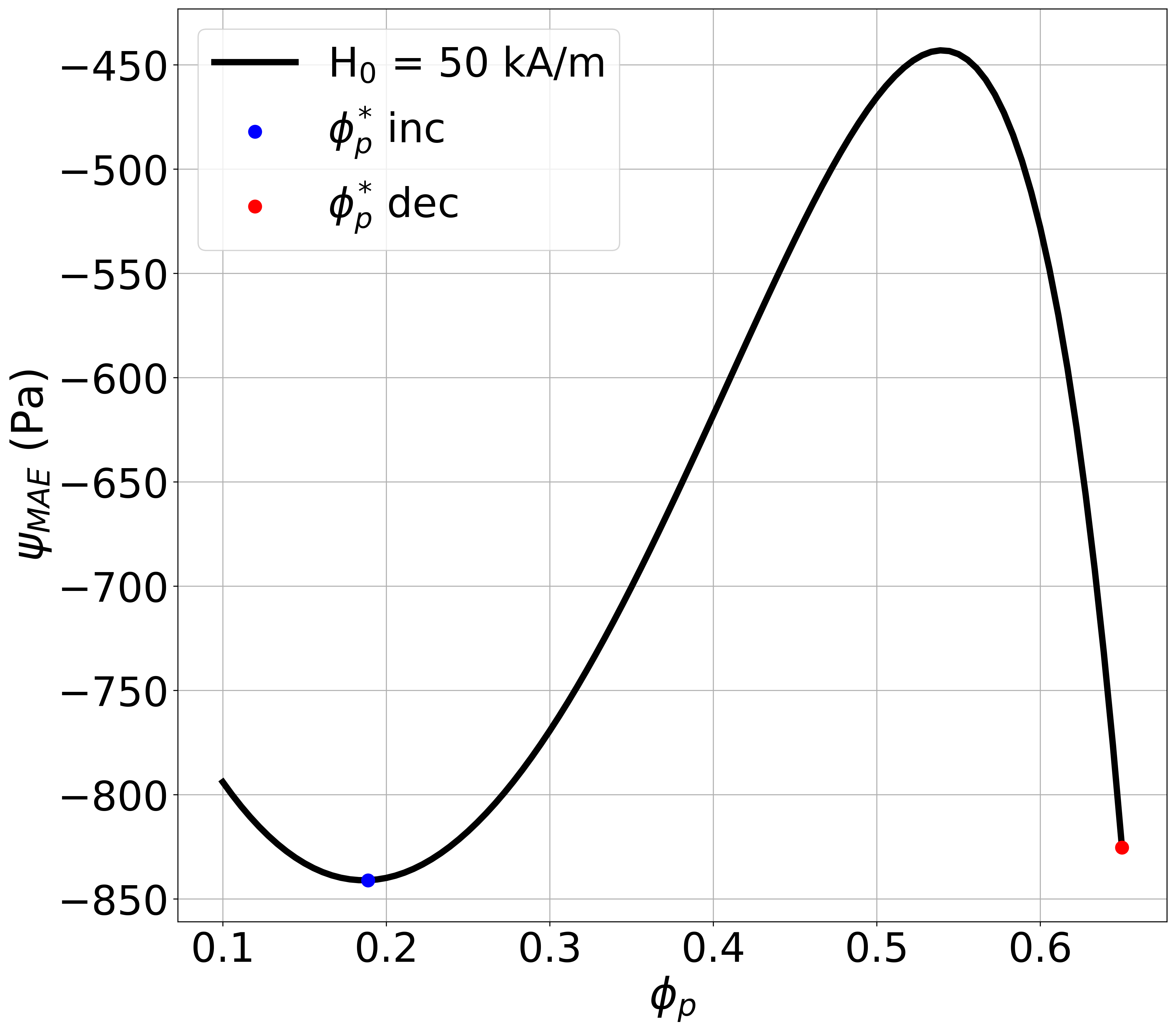}
        \caption{}
        \label{fig:c}
    \end{subfigure}
    \hfill
    \begin{subfigure}{0.4\textwidth}
        \centering
        \includegraphics[width=\linewidth]{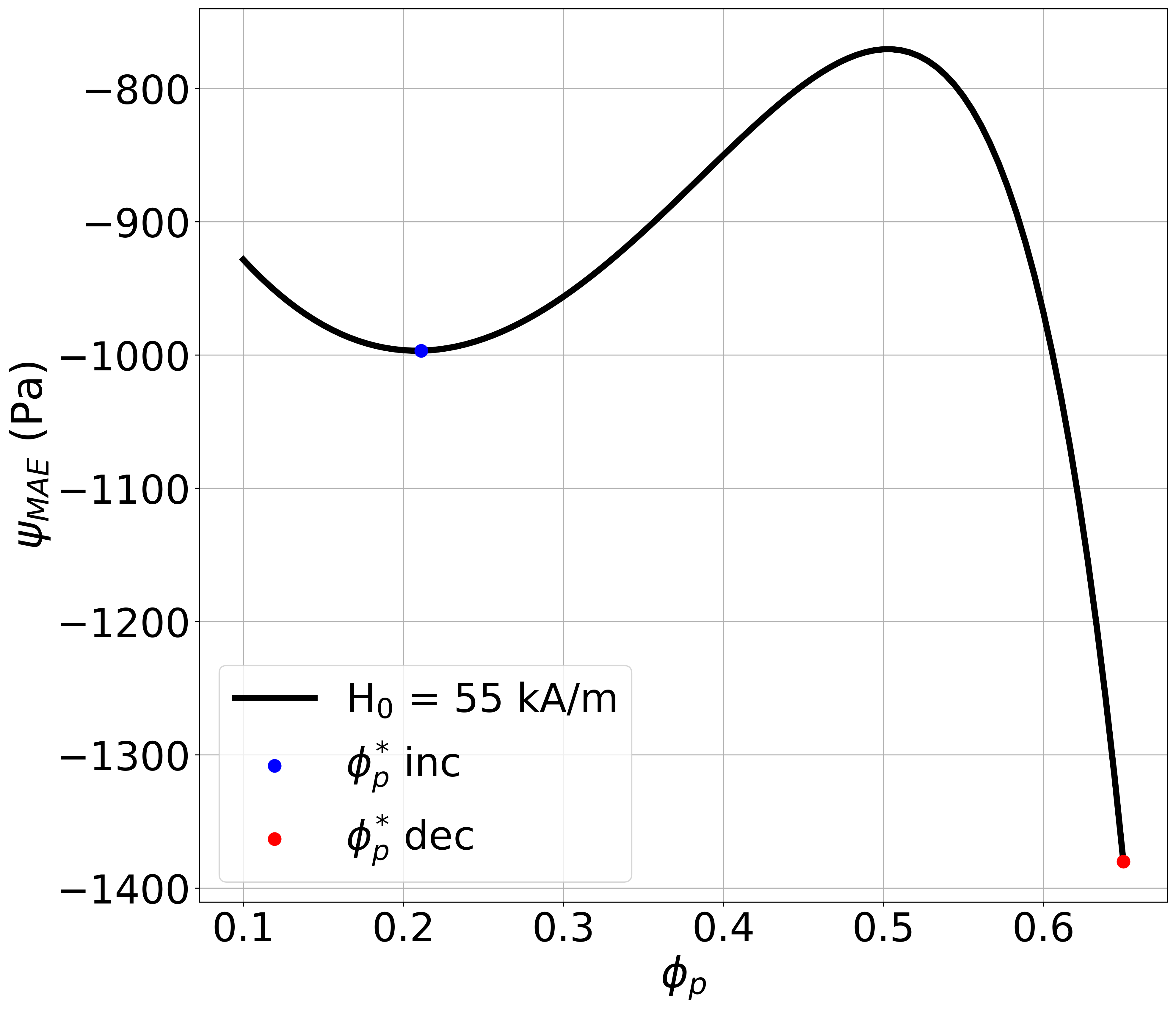}
        \caption{}
        \label{fig:d}
    \end{subfigure}

    \vspace{0.15cm}

    \begin{subfigure}{0.4\textwidth}
        \centering
        \includegraphics[width=\linewidth]{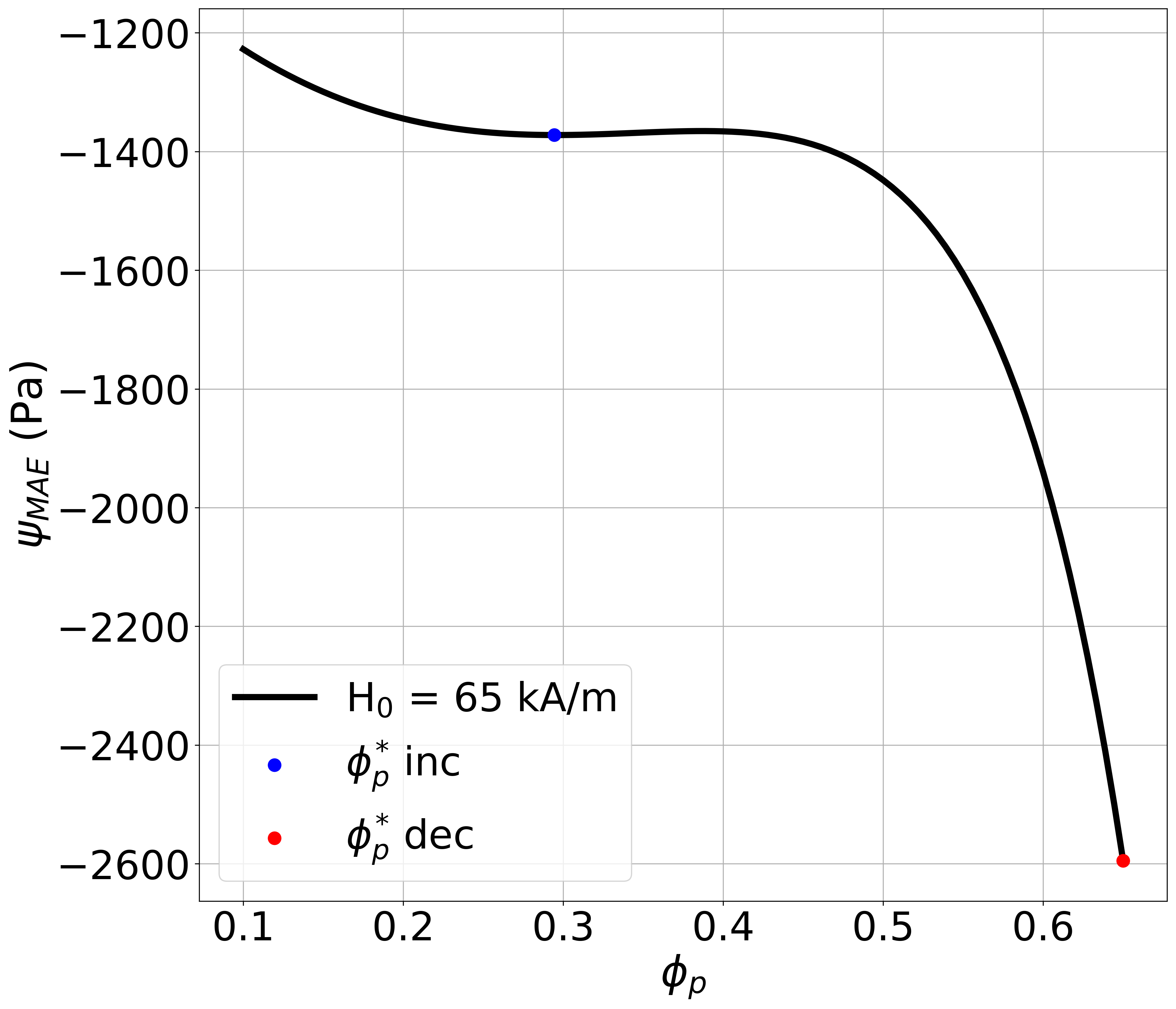}
        \caption{}
        \label{fig:e}
    \end{subfigure}
    \hfill
    \begin{subfigure}{0.4\textwidth}
        \centering
        \includegraphics[width=\linewidth]{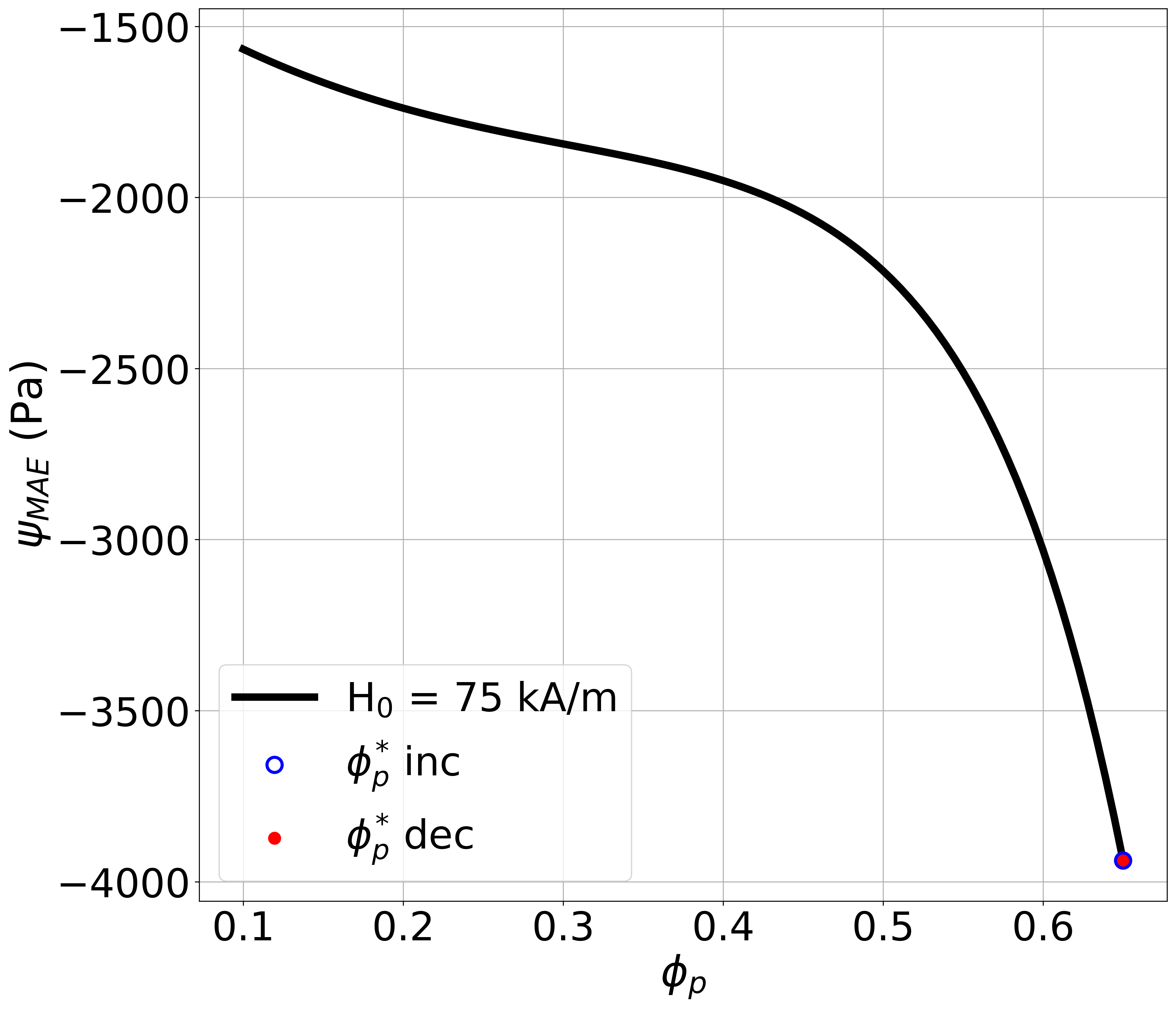}
        \caption{}
        \label{fig:f}
    \end{subfigure}

    \vspace{0.15cm}

    \caption{\label{fig:Fig4} Computation of equilibrium local volume fraction $\phip^{*}$ for increasing (open and filled blue circles) and decreasing (filled red circles) magnetic field. The parameters are $\phi = 0.1$, $\chi = 100$, $k_m = 5$, $G_{\mathrm{iso}} = 30\, \text{kPa}$, $f_{macro} = 0.1$. Near-field effects are included.}
\end{figure}

Assuming the Fröhlich-Kennelly saturation function, the magnetization is related to the local magnetic field by:
\begin{equation}
    M = \frac{\chi H}{1 + \frac{\chi H}{M_{\text{sat}}}}
\end{equation}
where $M_{\text{sat}}$ denotes the saturation magnetization. For a given applied field $H_0$, the particle volume fraction inside the columnar structures, $\phip$, is determined from minimization of the total energy. The corresponding equilibrium value is denoted by $\phip^{*}$. To compute $\phip^{*}$ numerically, the total energy is evaluated over a range of $\phip$ values at fixed $H_0$, starting from $\phip=\phi$ and increasing up to $\phi_{\text{max}}=0.65$, consistent with the analysis in the linear magnetization regime. The equilibrium value $\phip^{*}$ is subsequently used to compute the total energy and the induced magnetization $M$. The saturation magnetization is taken from experiments~\cite{Roghani2023} as $M_{\text{sat}}=1500 \text{kA/m}$ and is used in the Fröhlich–Kennelly relation. The model depends on several parameters, including the total particle volume fraction $\phi$, the initial susceptibility $\chi$, the sample aspect ratio $\gamma$, the micromechanical constant $k_m$, and the shear modulus of the polymer matrix $G_{\mathrm{iso}}$. In the present calculations, $\chi=100$ and $G_{\mathrm{iso}} = 30$ kPa are kept fixed, with $G_{\mathrm{iso}}$ chosen to represent a realistic elastomer matrix \cite{Roghani2025}. All parameter values used are summarized in Fig.~\ref{fig:Fig4}.

Fig.~\ref{fig:Fig4} illustrates the evolution of the energy landscape for increasing and decreasing magnetic field. At low fields, the energy exhibits a single minimum (Fig.~5(a)), which defines $\phip^{*}$. At intermediate fields (Figs.~5(b-d)), a second extremum appears, separated by an energy barrier. Since the particles are too large for Brownian motion, thermally activated barrier crossing is neglected, and the system remains trapped in the metastable minimum. Upon further increasing the field, the energy landscape develops an extended flat region (plateau), as shown in Fig.~5(e), representing the state immediately before the discontinuous transition to the densely packed configuration. In this case, $\phip^{*}$ is obtained by averaging the values of $\phip$ within the plateau region. At higher fields, the plateau disappears and $\phip^{*}$ jumps to the high-density minimum corresponding to tightly packed columns (Fig.~5(f)).
When decreasing the field from this state, the system initially remains trapped in the densely packed configuration, (Fig.~5(f)). With further reduction of $H_0$, the energy barrier prevents an immediate transition back to the low-density minimum, and the system remains on the metastable branch (Figs.~5(b–e)). Only at sufficiently low fields (Fig.~5(a)) does the system return to the dispersed state, where both branches coincide again. Overall, the particle distribution follows different equilibrium paths for increasing and decreasing field, resulting in hysteresis.
 
\begin{figure}[htbp]
\centering
  \includegraphics[width=0.7\textwidth]{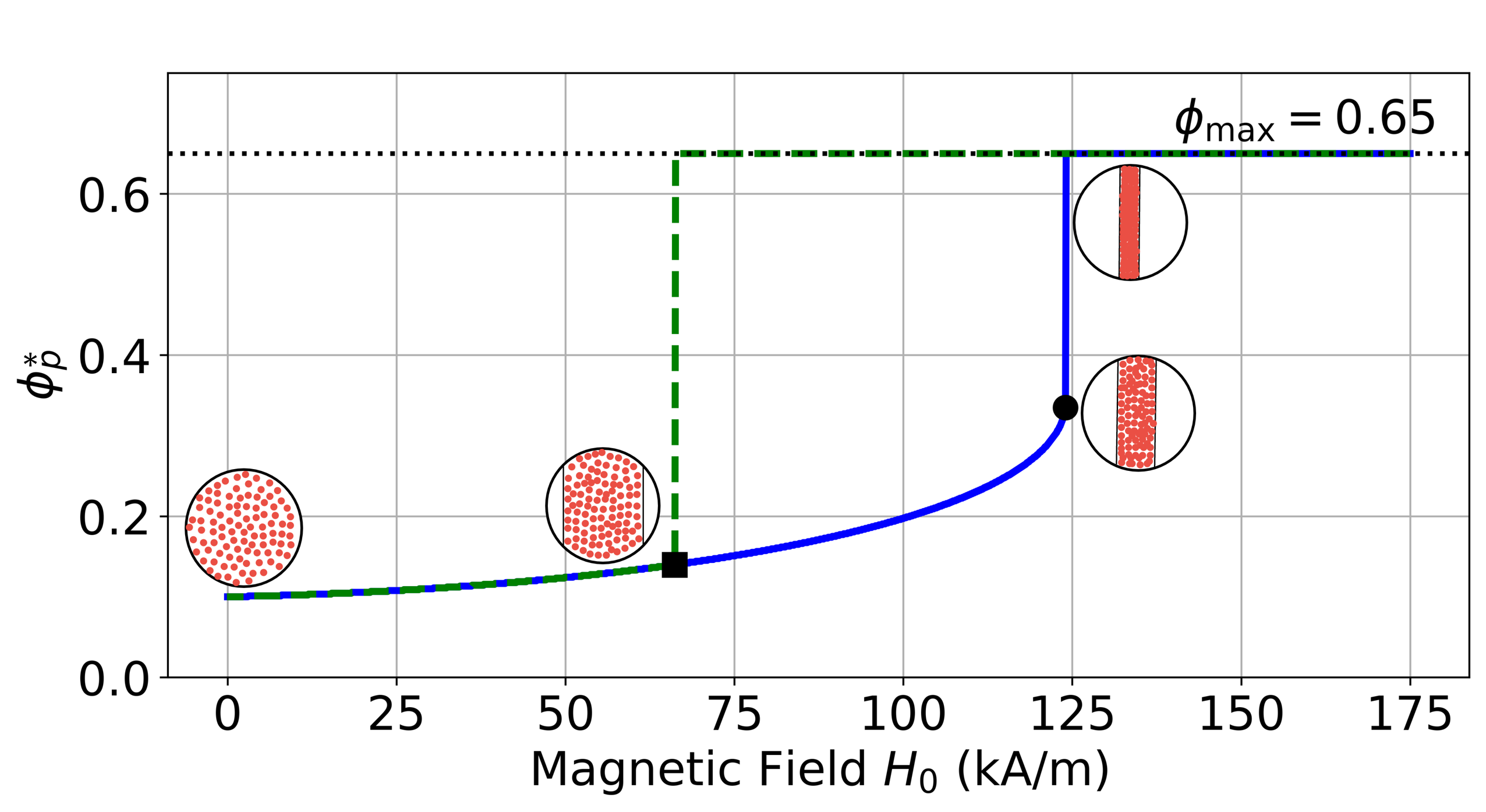}
  \caption{
  Microstructure evolution quantified by the equilibrium local particle volume fraction $\phip^{*}$ as a function of the applied field $H_0$. Inserts schematically illustrate the corresponding particle arrangements. Black circle and square indicate the discontinuous transitions on the increasing and decreasing field branches.}
  \label{fgr:Fig5}
\end{figure}

\begin{figure}[htbp]
    \centering

    \begin{subfigure}[b]{0.47\textwidth}
        \centering
        \includegraphics[width=\textwidth]{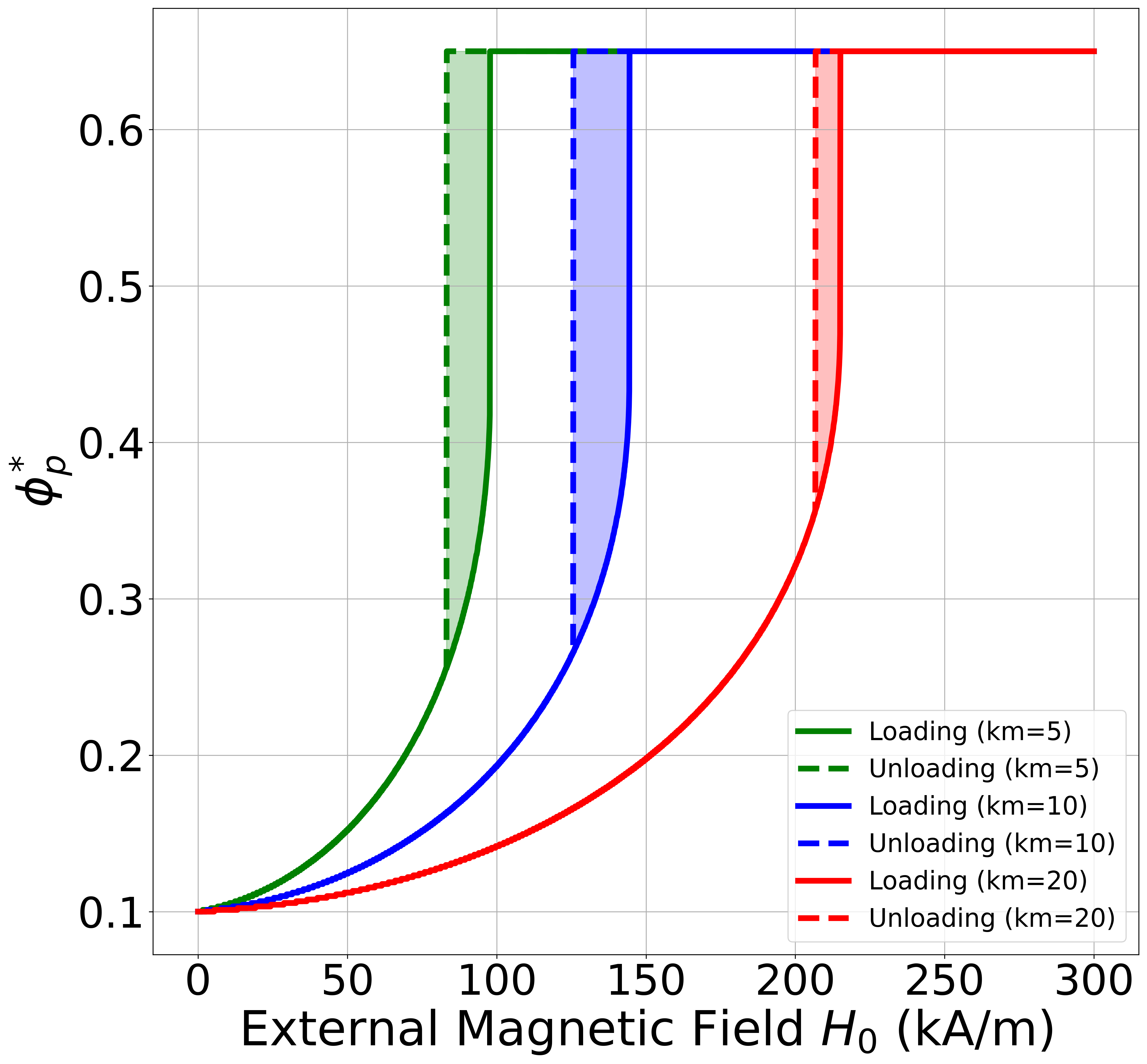}
        \caption{}
        \label{fig:hyst_noNFE_a}
    \end{subfigure}
    \hfill
    \begin{subfigure}[b]{0.47\textwidth}
        \centering
        \includegraphics[width=\textwidth]{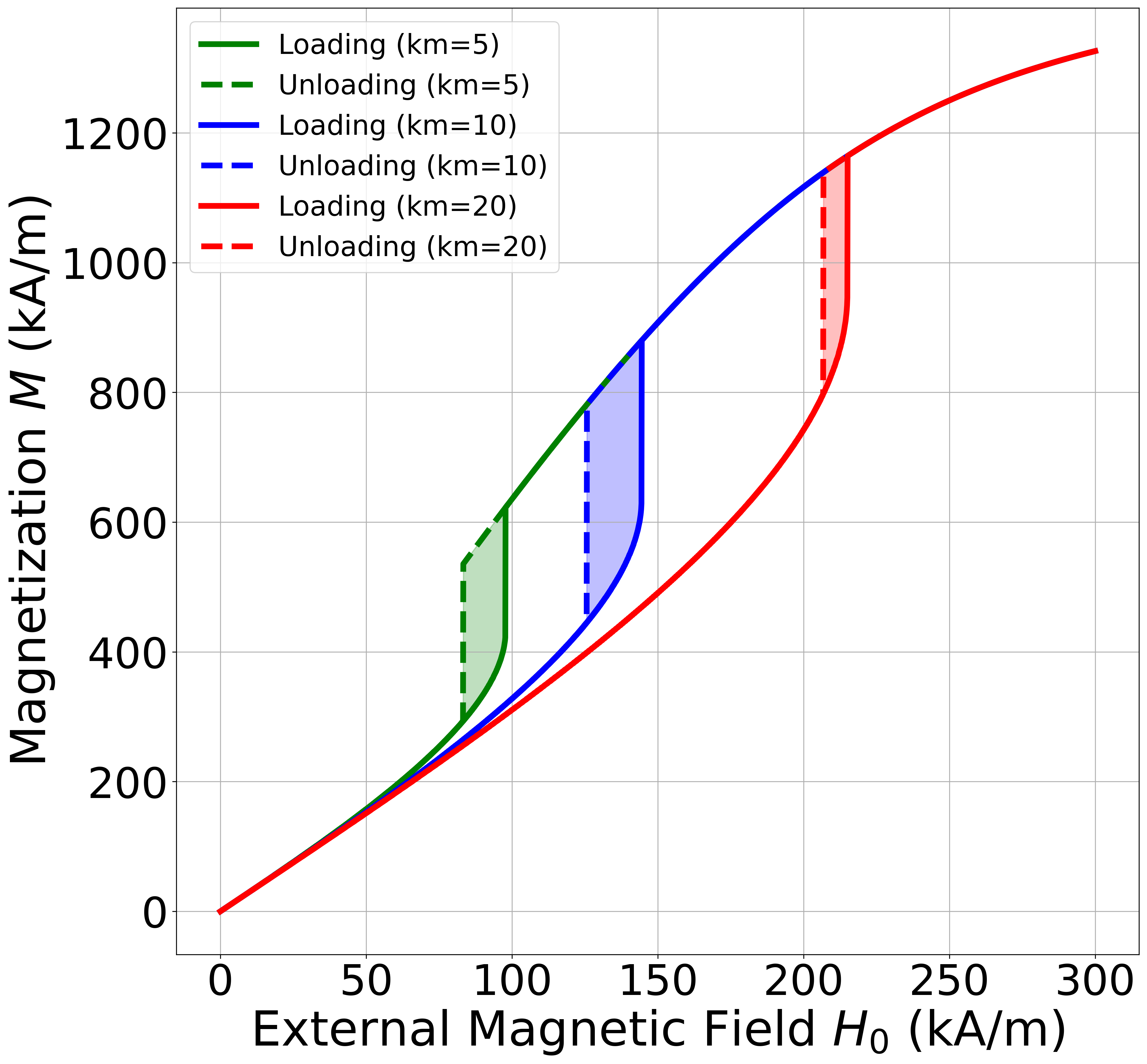}
        \caption{}
        \label{fig:hyst_noNFE_b}
    \end{subfigure}

    \vspace{0.4cm}

    \begin{subfigure}[b]{0.47\textwidth}
        \centering
        \includegraphics[width=\textwidth]{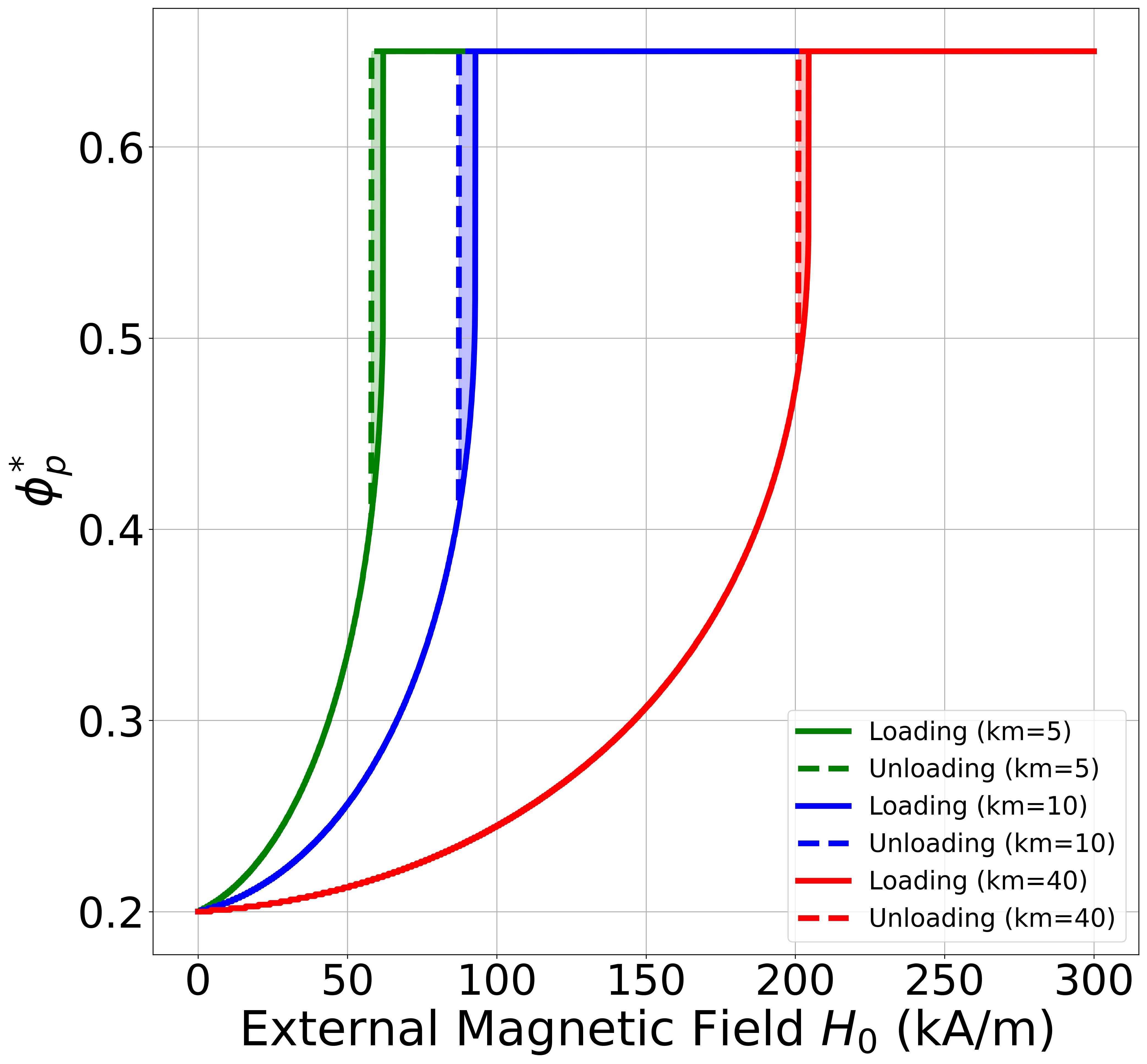}
        \caption{}
        \label{fig:hyst_noNFE_c}
    \end{subfigure}
    \hfill
    \begin{subfigure}[b]{0.47\textwidth}
        \centering
        \includegraphics[width=\textwidth]{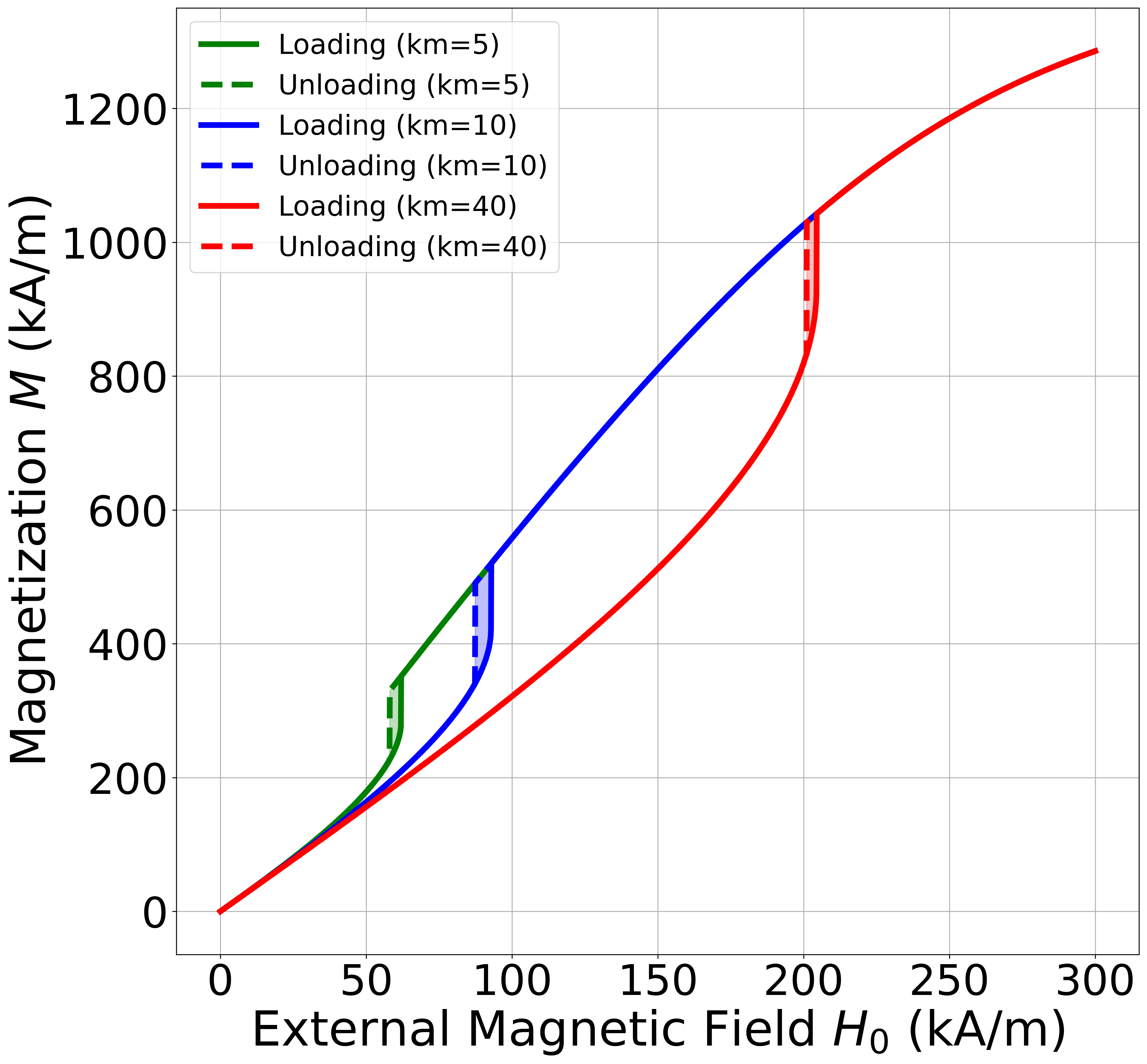}
        \caption{}
        \label{fig:hyst_noNFE_d}
    \end{subfigure}

    \caption{Effect of the micromechanical constant $k_m$ on the hysteresis loop width without NFE. Panels (a,c) show the evolution of the local particle volume fraction $\phip^{*}$, while panels (b,d) show the corresponding magnetization. Results are shown for $\phi = 0.10$ (a,b) and $\phi = 0.20$ (c,d).}
    \label{fig:hyst_noNFE}
\end{figure}

\begin{figure}[htbp]
    \centering

    \begin{subfigure}[b]{0.47\textwidth}
        \centering
        \includegraphics[width=\textwidth]{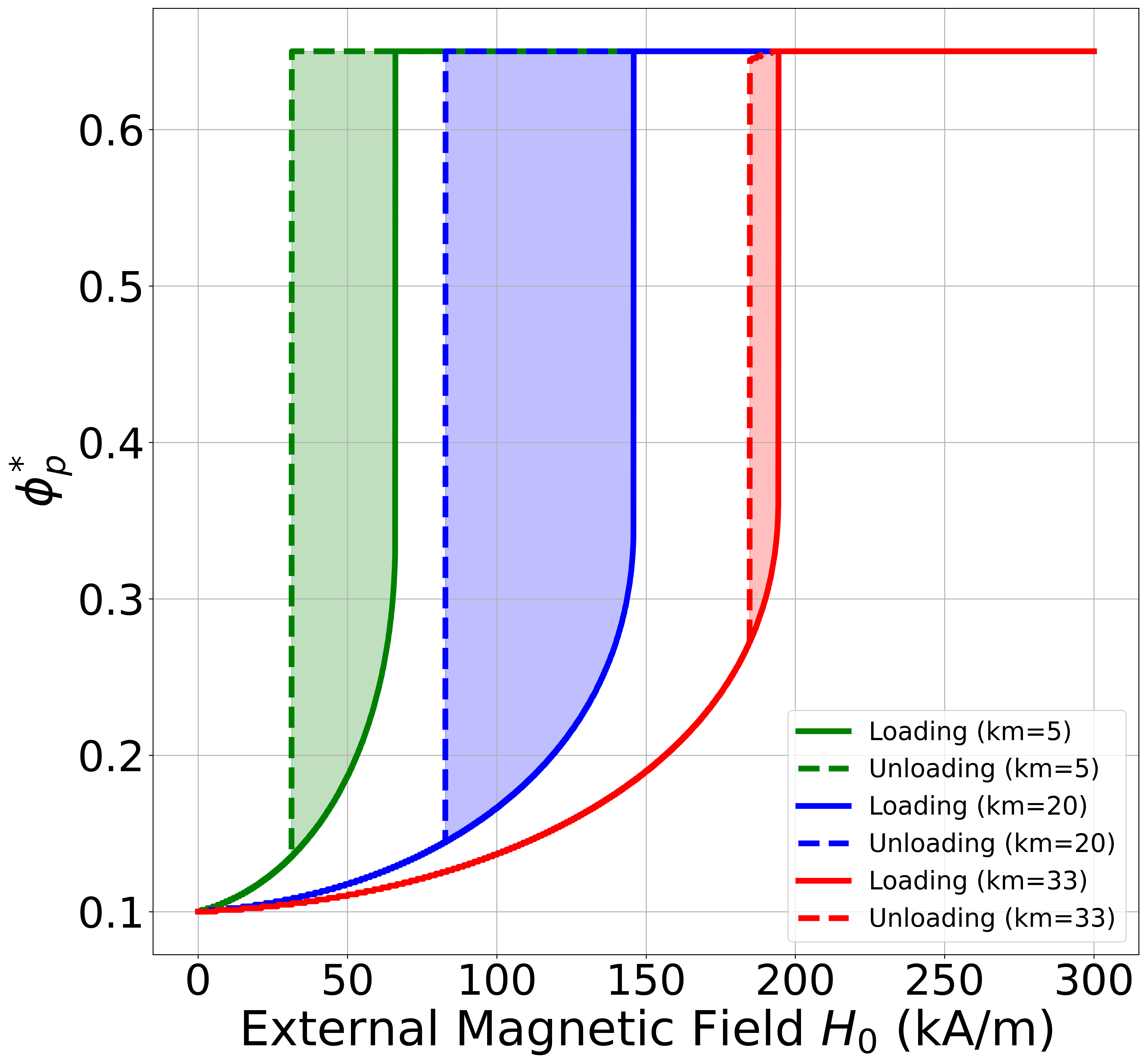}
        \caption{}
        \label{fig:hyst_NFE_a}
    \end{subfigure}
    \hfill
    \begin{subfigure}[b]{0.47\textwidth}
        \centering
        \includegraphics[width=\textwidth]{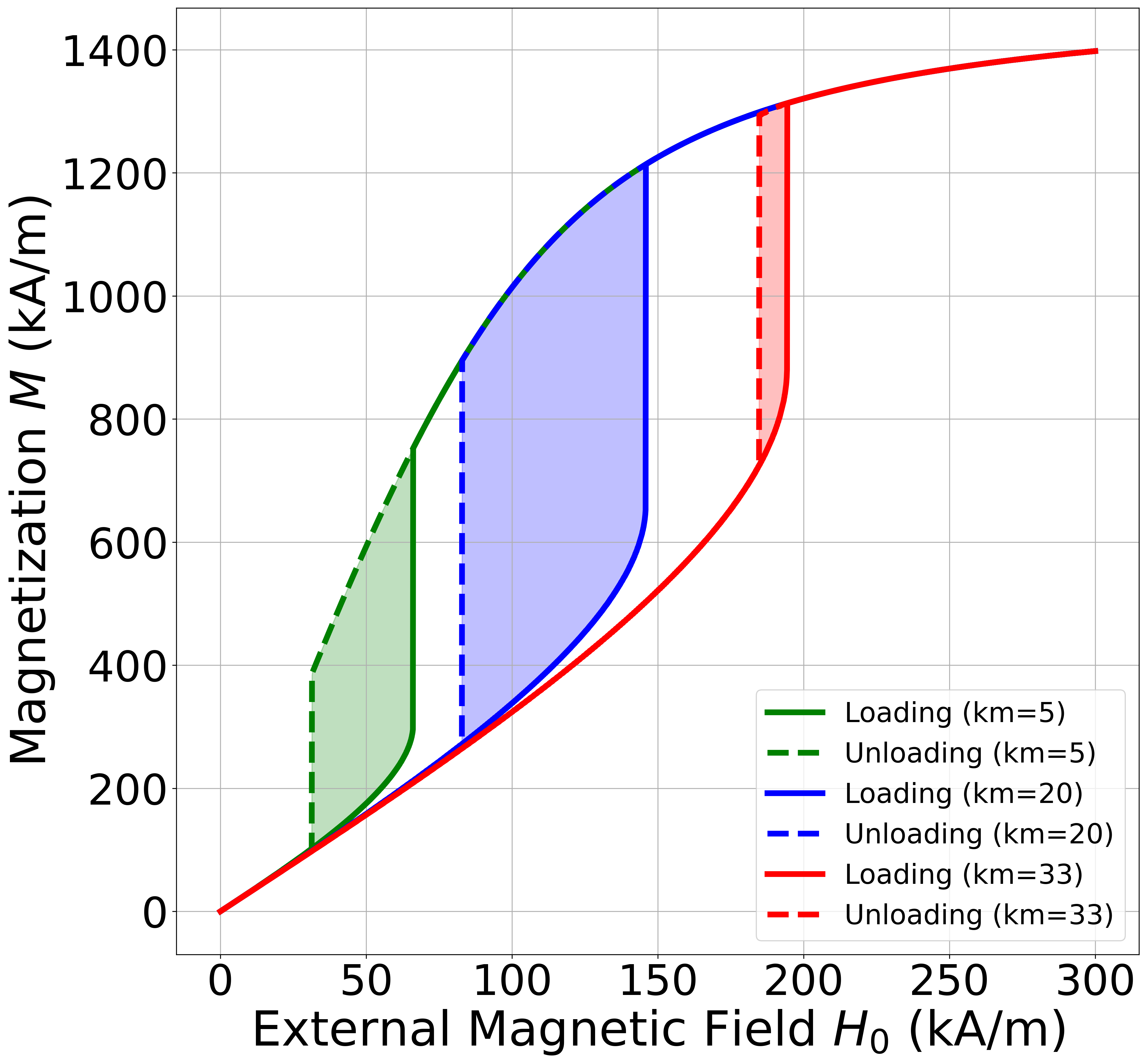}
        \caption{}
        \label{fig:hyst_NFE_b}
    \end{subfigure}

    \vspace{0.4cm}

    \begin{subfigure}[b]{0.47\textwidth}
        \centering
        \includegraphics[width=\textwidth]{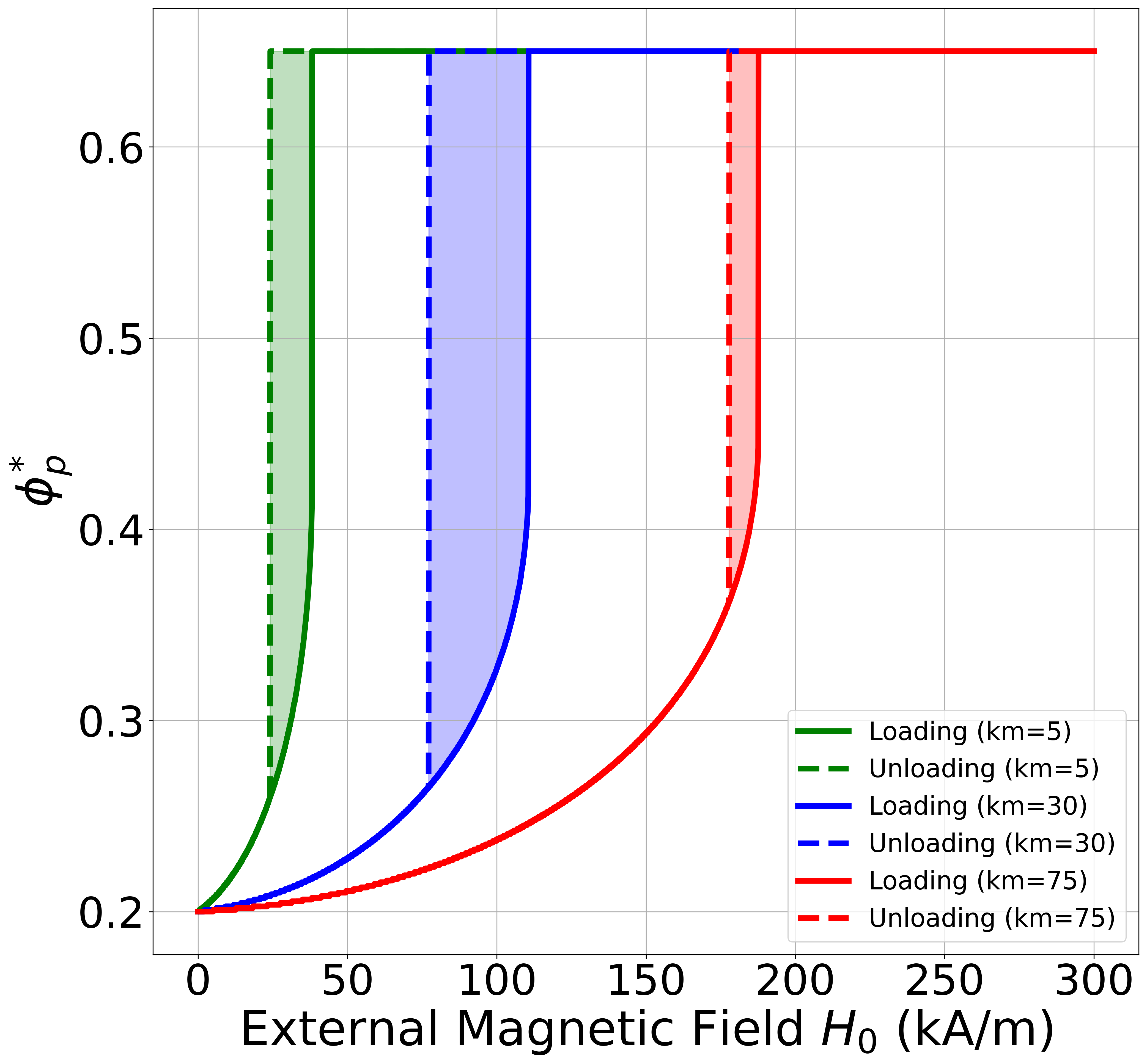}
        \caption{}
        \label{fig:hyst_NFE_c}
    \end{subfigure}
    \hfill
    \begin{subfigure}[b]{0.47\textwidth}
        \centering
        \includegraphics[width=\textwidth]{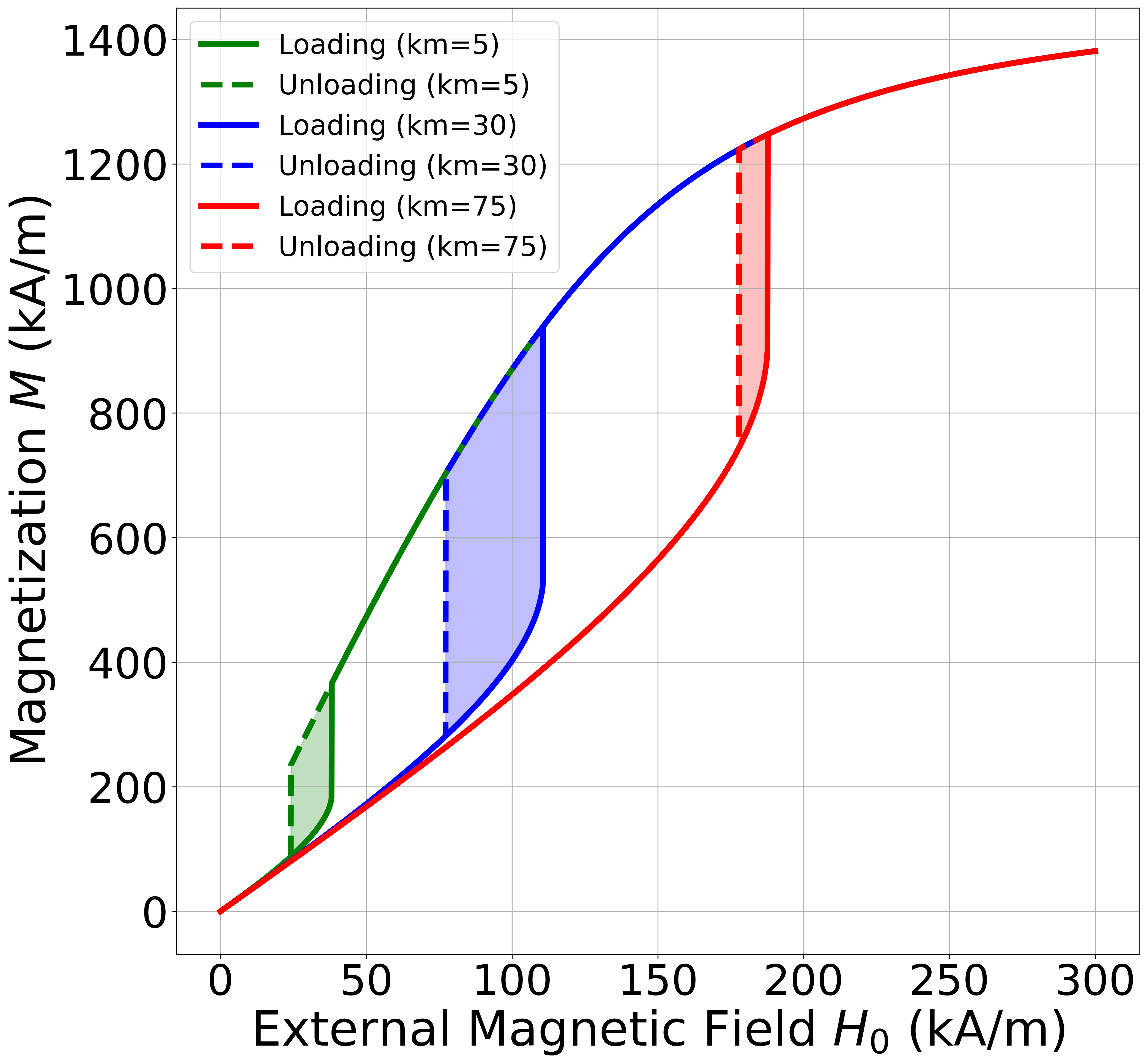}
        \caption{}
        \label{fig:hyst_NFE_d}
    \end{subfigure}

    \caption{Effect of the micromechanical constant $k_m$ on the hysteresis loop width with NFE. Panels (a,c) show the evolution of the local particle volume fraction $\phip^{*}$, while panels (b,d) show the corresponding magnetization. Results are shown for $\phi = 0.10$ (a,b) and $\phi = 0.20$ (c,d).}
    \label{fig:hyst_NFE}
\end{figure}

\begin{figure}[htbp]
\centering
  \includegraphics[width=0.7\textwidth]{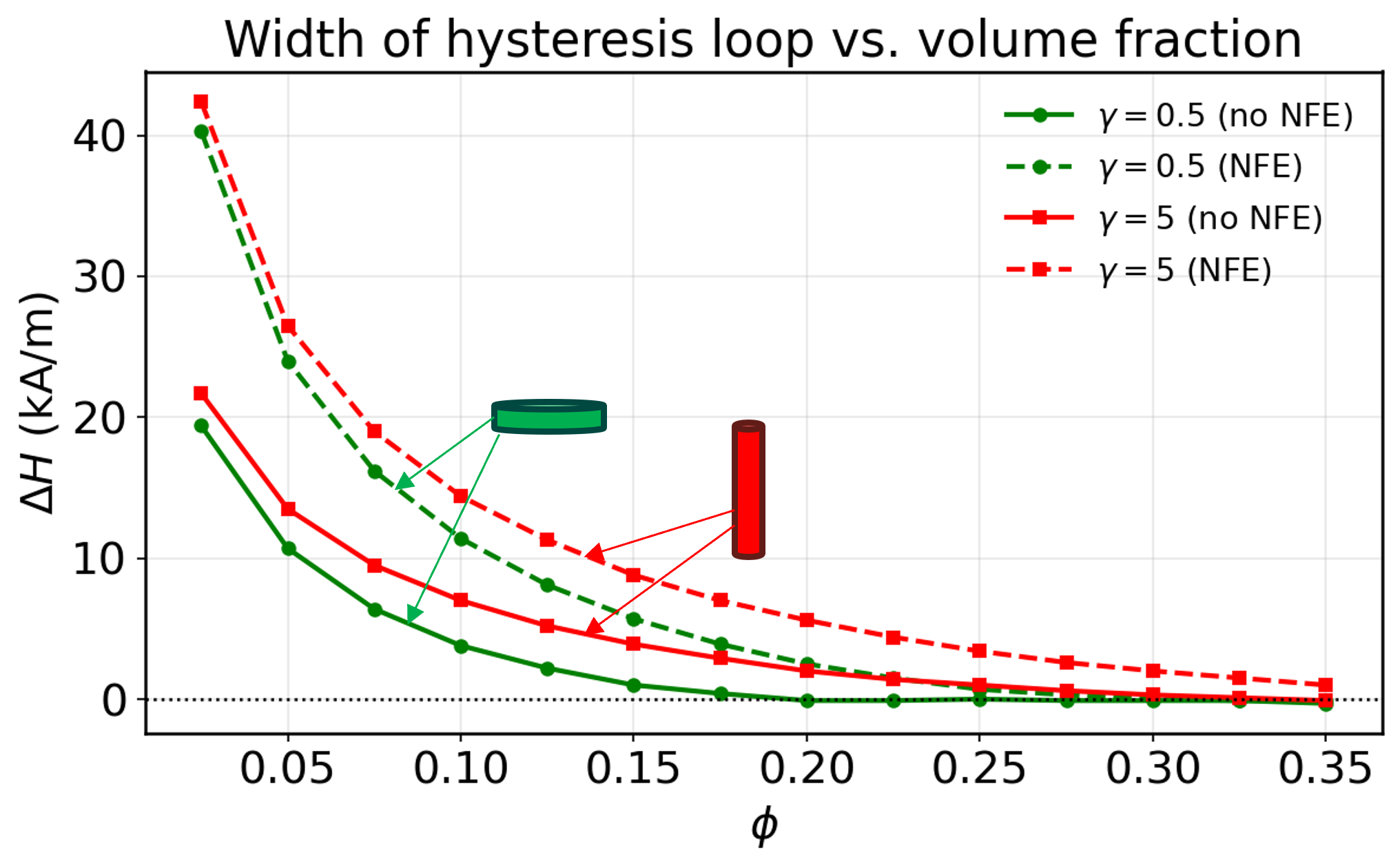}
  \caption{\label{fgr:Hystphi} Effect of total volume fraction on the width of the hysteresis loop for different shapes of the sample.}
\end{figure}

The internal magnetic field $H$ is first evaluated using a pure dipole–dipole interaction model. The equilibrium local particle volume fraction $\phip^{*}$ increases with increasing external field $H_0$ and decreases as $H_0$ is reduced, consistent with the underlying microstructural evolution. This dependence is schematically shown in Fig.~\ref{fgr:Fig5}. At $H_0=0$, the particles are uniformly distributed within the elastomer matrix, corresponding to an isotropic state with $\phip^{*}=\phi$ (first insert). With increasing field, the particles approach each other and progressively form columnar structures (second and third inserts), leading to an increase in $\phip^{*}$. At a critical field $H_0^{*\uparrow}$, a discontinuous jump occurs (marked by the black circle on the increasing branch), indicating the transition to a densely packed columnar state with $\phip^{*}=\phi_\text{max}$. Beyond this point, the MAE is strongly anisotropic (fourth insert), and further increases in $H_0$ do not result in an additional increase of the particle volume fraction within the columns. Upon decreasing the field, the strongly anisotropic state persists until the second critical field $H_0^{*\downarrow}$ is reached (black square on the decreasing branch), at which point the energy barrier disappears and the system jumps back towards a more dispersed configuration. Further reduction of $H_0$ leads to continuous dispersion of the columns, and the isotropic state is recovered when the field is removed. Overall, Fig.~\ref{fgr:Fig5} illustrates how microstructure evolution, quantified by $\phip^{*}$, gives rise to magnetic hysteresis.
Fig.~\ref{fig:hyst_noNFE} presents the width of the hysteresis loops for different values of $k_m$ and $\phi$ in the absence of NFE. The parameter $k_m$ characterizes the degree of microstructural mobility. As discussed in the previous section, under the assumption of linear magnetization, larger values of $k_m$ result in wider hysteresis loops, as illustrated in Fig.~4(b). In this case, the increase in loop width is unbounded. In contrast, when magnetization saturation is taken into account, the loop width exhibits a non-monotonic behavior. Specifically, it increases up to a threshold value, beyond which it decreases and eventually vanishes. This behavior arises because the magnetization initially follows a linear regime, then transitions through a nonlinear region, and finally approaches saturation. Figs.~\ref{fig:hyst_noNFE} and~\ref{fig:hyst_NFE} illustrate this trend. In particular, Fig.~\ref{fig:hyst_noNFE} shows the influence of $k_m$ within the pure dipole–dipole interaction model. Two representative cases corresponding to different particle volume fractions are considered. For $\phi = 0.10$, the hysteresis loop width $\Delta H$ is 14.5~kA/m at $k_m = 5$. As $k_m$ increases to 10, $\Delta H$ increases to 18.8~kA/m. Beyond this point, a threshold is reached, after which $\Delta H$ decreases with further increases in $k_m$. To further examine this limiting behavior, the equilibrium local particle volume fraction $\phip^{*}$ is plotted as a function of $H_0$ in Figs. 7(a) and 7(b). Notably, the evolution of $\phip^{*}$ with respect to $H_0$ closely resembles the magnetization curves. The observed non-monotonic behavior is attributed to the nonlinear magnetization response.
The introduction of NFE significantly affects the hysteresis loop width, as it enables a more accurate representation of higher-order interactions between magnetized particles compared to the pure dipole–dipole model. Consequently, wider hysteresis loops are observed when NFE is added, as evidenced by the comparison between Figs.~\ref{fig:hyst_noNFE} and~\ref{fig:hyst_NFE}. Figs. 8(a) and 8(b) show the evolution of the equilibrium local particle volume fraction and magnetization considering NFE for $\phi = 0.10$. For $k_m = 5$, the loop width $\Delta H$ is 34.7~kA/m, which is significantly larger than that obtained for the pure dipole–dipole model ($\Delta H = 14.5$~kA/m). With increasing $k_m$, the loop width follows a trend similar to that observed in the dipole–dipole case. As the particle volume fraction increases to $\phi = 0.20$, the hysteresis loops become narrower, as shown in Fig. 7(d), where $\Delta H$ decreases to 3.9~kA/m for $k_m = 5$. A similar reduction in loop width is observed when NFE is incorporated, as illustrated in Fig. 8(d).
Next, the influence of particle volume fraction on the hysteresis behavior is examined. A comparison of hysteresis loops for different values of $\phi$ is presented in Fig.~\ref{fgr:Hystphi}, where the loops are significantly wider for MAEs with low particle volume fractions compared to those with higher $\phi$. This behavior is attributed to the greater extent of microstructural evolution in systems with lower particle concentrations, which enhances the hysteretic response. An important observation is that the hysteresis loops vanish at $\phi = 0.35$, indicating that particle mobility becomes strongly restricted and that particles can no longer rearrange along the direction of the applied magnetic field. Consequently, highly filled MAEs do not exhibit hysteresis.
Finally, the effect of sample shape on the hysteresis behavior is considered. The parameter governing shape effects is $f_{macro}$, which, as defined in Eqs.~\ref{eq_fmacro} and~\ref{eq_demag_shape}, depends on the sample aspect ratio $\gamma$. Two representative geometries are examined: $\gamma = 0.5$, corresponding to a disc-like shape, and $\gamma = 5$, corresponding to a long cylindrical geometry. The results indicate that the hysteresis loop is narrower for the flat disc than for elongated cylinder. This difference arises from the variation in demagnetization effects associated with sample geometry. The model therefore captures the influence of shape on the hysteresis response. Consistent with previous observations, the inclusion of NFE leads to wider hysteresis loops.
\section{Discussion and conclusion}
A comprehensive model has been developed to capture the magnetic hysteresis behavior of magneto-active elastomers (MAEs) by coupling magnetic and microstructural effects. The results demonstrate that hysteresis in MAEs with soft magnetic fillers originates from the different configurations of the microstructure under increasing and decreasing magnetic fields. The influence of the micromechanical constant is found to depend on the magnetization regime. In the case of linear magnetization, increasing the micromechanical constant leads to wider hysteresis loops. In contrast, for nonlinear (saturation) magnetization, the loop width initially increases with the micromechanical constant, reaches a maximum, and subsequently decreases beyond a threshold value. Both magnetization regimes exhibit similar trends with respect to variations in particle volume fraction: MAEs with low filler content show pronounced hysteresis, whereas highly filled systems exhibit nearly reversible behavior due to restricted particle mobility. The mean field model also captures the shape effect on hysteresis. Disc shaped samples exhibit narrower loops compared to the long cylindrical samples. The model effectively describes the key microstructural transitions, including plateau formation, which corresponds to the threshold for abrupt structural rearrangements, and the transition from densely packed columns to more dispersed configurations. These features are clearly reflected in the energy landscape for both magnetization regimes, demonstrating the ability of the model to identify critical transition points associated with hysteresis behavior. The incorporation of NFE enhances the accuracy of the model by accounting for higher-order particle interactions, resulting in broader and more realistic hysteresis loops. This work presents the first ever approximation of NFE for systems comprising multiple interacting particles, providing a foundation for the development of more general formulations in future studies. The present mean-field framework assumes the formation of particle columns with uniform local volume fraction. However, experimental observations indicate that columns may exhibit spatial variations in local particle concentration \cite{Gundermann_2014}. Future work will therefore consider the formation of multiple columns with non-uniform volume fractions. In addition, FEA will be employed to better capture the averaging effects associated with such heterogeneous structures. Overall, the proposed framework establishes a direct link between microscale particle rearrangements and macroscale magnetic response. The ultimate objective is to validate and calibrate the model against experimental data, enabling accurate prediction of hysteresis behavior in MAEs.

\appendix*
\section{Approximation of the near-field effect}
Here, we introduce an approximation for the factor $f_{\text{NFE}}$ to account for near-field effects (NFE) when particles approach each other closely. Our formulation is partially inspired by the ideas proposed by Yaremchuk et al.~\cite{Yaremchuk_2024}, who derived an approximate treatment of NFE in many-body systems of magnetizable spheres. Their method is based on a compact expression for the magnetic energy of two linearly magnetizable particles in an external field $H_0$, as suggested by Biller et al.~\cite{Biller2014}. Using the known magnetic energy of the two-particle system, Yaremchuk et al.~\cite{Yaremchuk_2024} construct a self-consistent approximation for many-body systems, analogous to earlier approaches for pure dipole interactions~\cite{Ivaneyko_2014}. This construction assumes that each particle experiences an identical local environment, such that all particles are magnetized in the same way and contribute equally to the total magnetic energy. In the symmetric two-particle configuration, each particle contributes exactly one half of the total interaction energy. Based on this single-particle contribution, a self-consistent description of the many-body problem can be formulated~\cite{Yaremchuk_2024}. Inspired by this concept, we propose an alternative approximation for NFE in many-body systems. A more general treatment of many-body interactions among magnetizable spheres will be addressed in future work. In the present study, we restrict ourselves to the case of identically magnetized particles.

The starting point is the magnetic energy of two interacting spheres in an external field $H_0$. In the linear magnetization regime, the energy is given by:
\begin{equation}
\label{eq_A1}
U_{\text{mag}}= -\frac{\mu_0}{2}\int_{V}\!d^{3}r\left(\vec{M}\cdot\vec{H}_0\right)=-\frac{\mu_0}{2}\vec{H}_0\cdot\int_{V}\!d^{3}r\,\vec{M}=-\frac{\mu_0}{2}v_{\text{p}}\vec{H}_0\cdot\sum_{i=1}^{2}\langle\vec{M}_i\rangle=-\mu_0v_{\text{p}}H_0\langle M_{\parallel}\rangle~.
\end{equation}
This follows from symmetry: the magnetization field $\vec{M}(\vec{r})$ inside one sphere is the mirrored image of that inside the other sphere. Consequently, $\langle\vec{M}_1\rangle=\langle\vec{M}_2\rangle=\langle\vec{M}\rangle$. The component of $\vec{M}$ parallel to $\vec{H}_0$ is denoted by $M_{\parallel}$. Only this component contributes to the magnetic energy due to the scalar product with $\vec{H}_0$. The average magnetization over a sphere of volume $v_{\text{p}}$ is defined as $\langle\vec{M}\rangle=\frac{1}{v_{\text{p}}}\int_{v_{\text{p}}}d^{3}r\,\,\vec{M}$, while the surrounding medium is assumed to be non-magnetizable, i.e., $M=0$ in $V \backslash v_{\text{p}}$.

According to Biller et al.~\cite{Biller2014}, the exact solution for two interacting spheres, including NFE, can be accurately represented by the following compact expression:
\begin{equation}
\label{eq_A_Biller}
U_{\text{mag}}= -\mu_0v_{\text{p}}\chi_{\text{eff}}H_0^{2}\left\lbrace 1+\frac{v_{\text{p}}\chi_{\text{eff}}}{4\pi}\sum_{k=3}^{7}\, \left(\frac{\chi_{\text{eff}}}{3}\right)^{p_k-2}\left( \frac{a^{k-3}A_k}{(r-aB_k)^k} + \frac{a^{k-3}C_k\cos^{2}\vartheta}{(r-aD_k)^k} \right)\right\rbrace ~.
\end{equation}
Here, $\chi_{\text{eff}}=3\chi/(3+\chi)$ is the effective magnetic susceptibility of the particle material and $a$ denotes the radius of sphere. The fitting parameters $A_k$, $B_k$, $C_k$, $D_k$ and $p_k$ are given in \cite{Biller2014}. The center-to-center distance between the spheres is $r$, and the angle $\vartheta$ specifies the orientation between the external field $\vec{H}_0$ and the interparticle vector $\vec{r}=r\vec{e}_r$. For brevity, we denote the function in curly brackets by $\mathcal{F}$. Comparing Eq.~\ref{eq_A_Biller} with the general result in Eq.~\ref{eq_A1}, we obtain
\begin{equation}
\label{eq_A_Biller_short}
\langle M_{\parallel}\rangle = \chi_{\text{eff}}H_0\mathcal{F}~. 
\end{equation}
Thus, $\mathcal{F}=\mathcal{F}(r,\cos\vartheta)$ represents the functional dependence of the self-consistent solution for the average magnetization along $\vec{H}_0$. 

Within the dipole approximation, the spheres are assumed to be homogeneously magnetized, i.e., $\vec{M}\stackrel{!}{=}\langle\vec{M}\rangle$ throughout the particle volume. The corresponding dipole moment is $\vec{m}=v_{\text{p}}\vec{M}$. In the following, we consider only the component of $\vec{m}$ parallel to the external field and neglect any perpendicular contributions. For two interacting dipoles aligned with $\vec{H}_0$, the self-consistent solution follows from:
\begin{equation}
\label{eq_A_dp}
M_{\parallel}=\chi\left(H_0-\frac{1}{3}M_{\parallel}+v_{\text{p}}g M_{\parallel}\right)=\chi_{\text{eff}}\left(H_0+v_{\text{p}}g M_{\parallel}\right)~,
\end{equation}
with
\begin{equation}
\label{eq_dp_op}
g=\frac{1}{4\pi}\frac{3\cos^{2}\vartheta-1}{r^{3}}~.
\end{equation}
Here, $g=g(r,\cos\vartheta)$ denotes the dipole interaction operator along $\vec{H}_0$. 
Solving Eq.~\ref{eq_A_dp} self-consistently for $M_{\parallel}$ yields:
\begin{equation}
\label{eq_A_dp_solve}
M_{\parallel}=\chi_{\text{eff}}H_0\left\lbrace\left(1-v_{\text{p}}\chi_{\text{eff}}\,g\right)^{-1}\right\rbrace~.
\end{equation}
Analogously to the exact solution, the term in the curly brackets represents the functional dependence of the self-consistent solution arising from the dipole interaction operator $g$, when only the component parallel to $\vec{H}_0$ is retained. 

Comparing Eqs.\ \ref{eq_A_Biller_short} and \ref{eq_A_dp_solve}, we can formally construct a 'full' interaction operator for two interacting spheres:
\begin{equation}
\label{eq_A_full_op}
\mathcal{G}_{\text{full}} = \frac{{\mathcal F}-1}{v_{\text{p}}\,\chi_{\text{eff}}\,{\mathcal F}}~,
\end{equation}
and write:
\begin{equation}
\label{eq_A_full_eq}
\langle M_{\parallel}\rangle = \chi_{\text{eff}}\left(H_0+v_{\text{p}}\mathcal{G}_{\text{full}}\langle M_{\parallel}\rangle\right)~.
\end{equation}
This provides a compact expression describing the full interaction between two magnetizable spheres. In the case of two particles, this formulation is exact. For the dipole operator in Eq.~\ref{eq_A_dp}, the expression can be directly extended to a system of $N$ interacting particles, since the dipole fields of homogeneously magnetized spheres superpose linearly. This results in a system of $N \times N$ linear equations to be solved:
\begin{equation}
\label{eq_A_manydp}
M_{\parallel_i} = \chi_{\text{eff}}\left(H_0+ v_{\text{p}}\sum_{j\neq i}^{N}g_{ij}\,M_{\parallel_j}\right)~.
\end{equation}
In contrast, for the full field operator in Eq.~\ref{eq_A_full_eq}, such a linear superposition is no longer strictly valid. The local magnetization inhomogeneities within individual particles, implicitly contained in $\mathcal{G}_{\text{full}}$, will generally differ when three or more particles interact.
Nevertheless, we regard this formulation as a first approximation for incorporating near-field effects in many-particle systems. In the present approach, the mutual interactions are summed first, and the self-consistent solution is subsequently obtained from the resulting system of equations. This constitutes the key difference compared to the approach proposed by Yaremchuk et al.~\cite{Yaremchuk_2024}.

In the present work, we do not explicitly resolve the individual particle positions, but describe the particle distribution in terms of continuous concentration fields. While this reduces the accuracy of microstructure-specific effects, it avoids the practically impossible task of determining all particle positions in an MAE sample on the local scale.
Representing the discrete particle arrangement by continuous probability distributions therefore provides a practical alternative and allows for a more compact formulation of the governing model equations.
According to our previous work \cite{Romeis2016, Chougale_2022}, the microstructure factor for a given continuous distribution of particle volume fractions $\Omega(\vec{r})$ is obtained from:
\begin{equation}
\label{eq_A_fmicro1}
f_{\text{micro}} = \frac{1}{\phi V_{\text{MS}}}\iint\limits_{V_{\text{MS}}}d^{3}r\,d^{3}r\,'\bigg\lbrace \Omega(\vec{r})\,\, g\!\left(\vec{r}-\vec{r}\,'\right)\, \Omega(\vec{r}\,')\,\,\Theta\!\left(\left|\vec{r}-\vec{r}\,'\right|-2a\right)\bigg\rbrace~.
\end{equation} 
Here, the integrals are taken over the volume $V_{\text{MS}}$ of a mesoscopic sphere with radius $R_{\text{MS}}$. This radius is chosen large enough that the particle distribution can be considered homogeneous at the boundary when viewed from the sphere center, yet small enough to neglect macroscopic shape effects. Typically, $R_{\text{MS}}$ is on the order of 10–20 times the average nearest-neighbor distance~\cite{Ivaneyko_2014}. The Heaviside step function $\Theta(x)$ ensures non-overlapping particles.

In this context, random (isotropic) particle distributions are described by a homogeneous field, $\Omega(\vec{r})=\phi$, equal to the total particle volume fraction. Setting $\vec{r}\to\vec{r}-\vec{r}'$, the first volume integral yields $V_{\text{MS}}$, and the remaining integral becomes
\begin{equation}
\label{eq_A_fmicro_dp}
f_{\text{micro}} = \phi\!\!\!\!\int\limits_{V_{\text{MS}}}d^{3}r\, g\!\left(\vec{r}\right)\,\Theta\!\left(r-2a\right) =\phi\!\!\int\limits_{2a}^{\infty}r^{2}dr\int\limits_0^{1}du\left\lbrace \frac{3u^{2}-1}{r^{3}} \right\rbrace =0~.
\end{equation} 
Hence, the microstructure effect vanishes for pure dipole interactions~\cite{Romeis2016, Romeis_2021}. In the evaluation, spherical coordinates were used. The integration over the azimuthal angle $\varphi$ yields a factor $2\pi$, since $g$ is independent of $\varphi$, and the substitution $u=\cos\vartheta$ was introduced.

Replacing the dipole interaction operator $g$ by the full operator $\mathcal{G}_{\text{full}}$, we can analogously obtain a first approximation of the microstructure factor including NFE for isotropic particle distributions. The corresponding integrals cannot be evaluated analytically due to the complex form of $\mathcal{F}$. Using numerical integration, we obtain:
\begin{equation}
\label{eq_A_fmicro_NFE}
f_{\text{micro}}^{\text{full}} \approx  \begin{cases} 0.104\phi ~~\text{for}~ \chi=100 & \\[2mm] 0.116\phi~~\text{for}~\chi=10000 \end{cases}.
\end{equation} 
According to \cite{Biller2014}, the expression in Eq.~\ref{eq_A_Biller} was constructed primarily for large values of $\chi$. For $\chi>100$, the result depends only weakly on $\chi$ and converges to $f_{\text{micro}}^{\text{full}} \approx 0.1163\phi$ in the limit $\chi\to\infty$. With the aim of finding a convenient approximation form for $f_{\text{micro}}^{\text{full}}$, we neglect this minor dependence on $\chi$ and approximate
$f_{\text{micro}}^{\text{full}} = \phi/9$.
Since the dipole contribution vanishes for isotropic particle distributions, we directly identify the near-field contribution in isotropic samples as $f_{\text{NFE}}=f_{\text{micro}}^{\text{full}}$.

In the present work, we not only consider isotropic particle distributions, but primarily focus on the formation of columnar-like particle arrangements. Also for such structures, we aim to keep the leading equations as simple as possible and to capture the dominant effects within a first-order approximation.
In \cite{Chougale_2022}, the microstructure contribution for columnar structures was significantly simplified by assuming that all particles possess the same magnetization and that the reference particle, used to evaluate the interaction, is located well inside the column.

\begin{figure}[ht]
\centering
  \includegraphics[height=8cm]{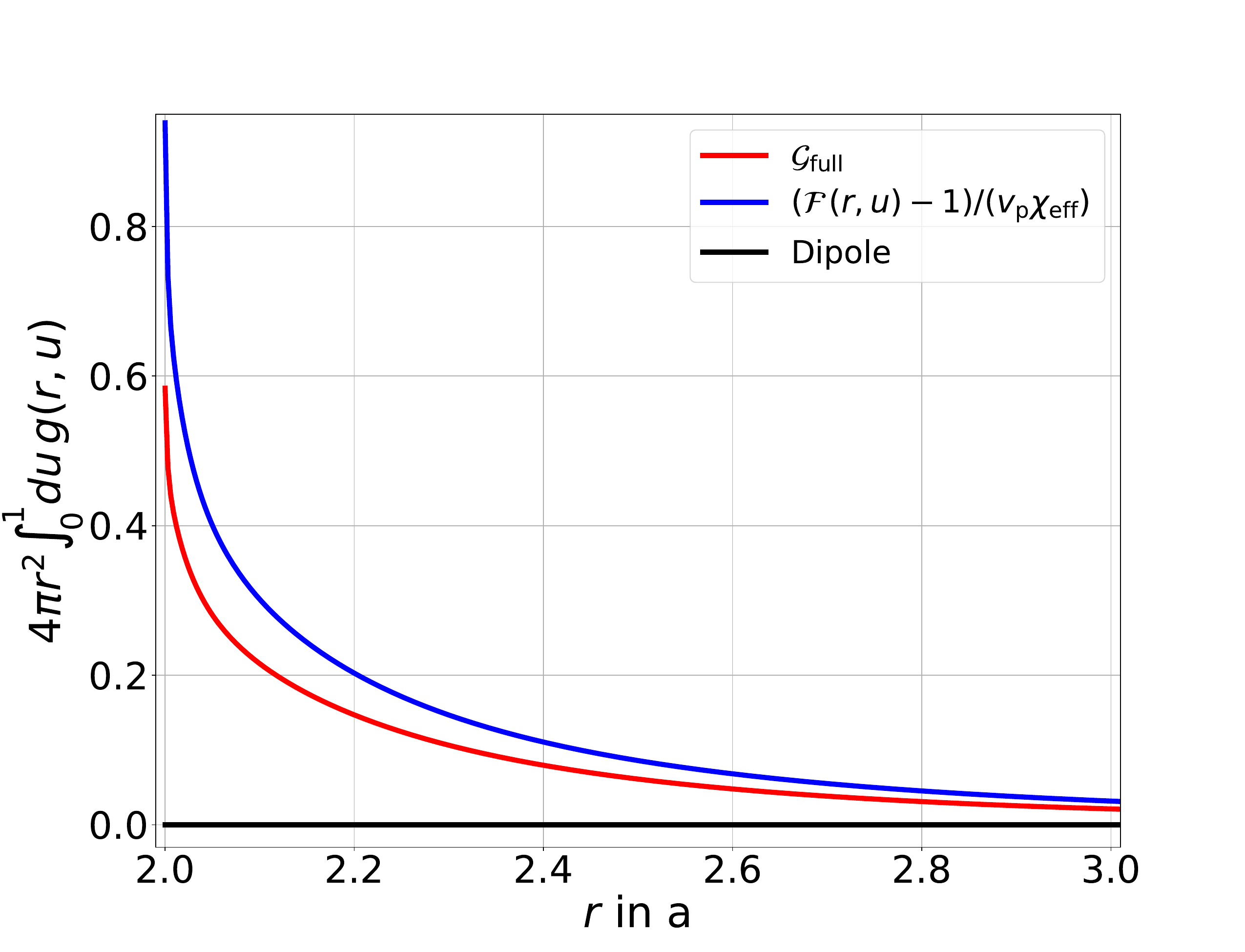}
  \caption{\label{fig_Appendix1} Comparing the different operators after integration over the azimuthal and polar angles.}
\end{figure}

In Fig.~\ref{fig_Appendix1}, we plot in red the radial dependence of the interaction operator $\mathcal{G}_{\text{full}}$ after integration over the azimuthal and polar angles. Here, $\chi=1000$ is assumed. For comparison, the corresponding operator proposed in \cite{Yaremchuk_2024}, which reads $\big(\mathcal{F}-1\big)/\big(v_{\text{p}}\chi_{\text{eff}}\big)$ in the present notation, is shown in blue.
The integral over this blue curve yields approximately $1/6$, about $50\%$ larger than the value obtained in our approach ($\sim 1/9$). The trivial zero result for the pure dipole operator is included in black. 

We observe that the NFEs decay rapidly with increasing distance $r$. For instance, at $r=3a$, corresponding to a gap of half a particle diameter, the NFEs are reduced by more than $95\%$ compared to the contact case $r=2a$. Therefore, when considering a reference particle located well inside a column, as in \cite{Chougale_2022}, it is a reasonable approximation to treat its surroundings as effectively isotropic with respect to NFE. In other words, near-field effects do not probe the column boundaries. 
To provide a compact representation of NFE in the governing equations, we therefore adopt:
\begin{equation}
\label{eq_A_f_NFE}
f_{\text{NFE}}=f_{\text{micro}}^{\text{full}}=\frac{\phi_{\text{p}}}{9}~.
\end{equation} 
This NFE approximation applies to both isotropic particle distributions ($\phi_{\text{p}}=\phi$) and columnar structures, where $\phi_{\text{p}}>\phi$.

\begin{acknowledgments}
The financial support provided by the DFG (German Research Foundation) within the framework of RTG 2430, research project 380321452, and the temporary position RO 6756/3-1 for Dirk Romeis is gratefully acknowledged. 
\end{acknowledgments}

\bibliography{references}

\end{document}